\documentclass[12pt,preprint]{aastex}
\usepackage{apjfonts,graphicx,graphics}
\usepackage{amsmath}
\usepackage{natbib}
\newcommand{\simless}
     {\ensuremath{\lower
3pt\hbox{$\rlap{\raise5pt\hbox{$\char'074$}}\mathchar"7218$}}}
\newcommand{\simgreat}
     {\ensuremath{\lower
3pt\hbox{$\rlap{\raise5pt\hbox{$\char'076$}}\mathchar"7218$}}}
\newcommand{\simgt}{\lower.5ex\hbox{$\; \buildrel > \over \sim \;$}}
\newcommand{\simlt}{\lower.5ex\hbox{$\; \buildrel < \over \sim \;$}}
\shorttitle{Magnetic Field Significance}
\shortauthors{Koch et al.}
\begin{document}
\title{The Importance of the Magnetic Field from an SMA-CSO-Combined Sample of Star-Forming Regions}
%

\author{
Patrick M. Koch\altaffilmark{1},
Ya-Wen Tang\altaffilmark{1}, 
Paul T. P. Ho\altaffilmark{1,2},
Qizhou Zhang\altaffilmark{2},
Josep Miquel Girart\altaffilmark{3},
Huei-Ru Vivien Chen\altaffilmark{4,1},
Pau Frau\altaffilmark{5},
Hua-Bai Li\altaffilmark{6},
Zhi-Yun Li\altaffilmark{7},
Hau-Yu Baobab Liu\altaffilmark{1},
Marco Padovani\altaffilmark{8,9},
Keping Qiu\altaffilmark{10} ,
Hsi-Wei Yen\altaffilmark{1},
How-Huan Chen\altaffilmark{2},
Tao-Chung Ching\altaffilmark{2,4},
Shih-Ping Lai\altaffilmark{4,1} and
Ramprasad Rao\altaffilmark{11}
}
\altaffiltext{1}{Academia Sinica, Institute of Astronomy and
 Astrophysics, Taipei, Taiwan}
\altaffiltext{2}{Harvard-Smithsonian Center for Astrophysics, 60
 Garden Street, Cambridge, MA 02138, USA}
\altaffiltext{3}{Institut de Ci\`{e}ncies de l'Espai, CSIC-IEEC, Campus UAB, 
Facultat de Ci\`{e}ncies, C5p 2, 08193 Bellaterra, Catalonia}
\altaffiltext{4}{Institute of Astronomy and Department of Physics, National Tsing Hua University, 
101 Section 2 Kuang Fu Road, Hsinchu 30013, Taiwan}
\altaffiltext{5}{Observatorio Astron\'{o}mico Nacional, Alfonso XII, 3 E-28014 Madrid, Spain}
\altaffiltext{6}{Department of Physics, The Chinese University of HongKong}
\altaffiltext{7}{Department of Astronomy, University of Virginia, P.O. Box 400325, Charlottesville, VA 22904, USA}
\altaffiltext{8}{Laboratoire Univers et Particules de Montpellier, UMR 5299 du CNRS, Universit\'e de Montpellier II, place E. Bataillon, cc072, 34095 Montpellier, France}
\altaffiltext{9}{INAF--Osservatorio Astrofisico di Arcetri, Largo E. Fermi 5, 50125 Firenze, Italy}
\altaffiltext{10}{School of Astronomy and Space Science, Nanjing University, 22 Hankou Road, Nanjiing 210093, China}
\altaffiltext{11}{Academia Sinica, Institute of Astronomy and Astrophysics, 645 N. Aohoku Place, Hilo, HI 96720, USA}
\email{pmkoch@asiaa.sinica.edu.tw}
%

%
\begin{abstract}

Submillimeter dust polarization measurements of a sample of 50 star-forming regions, observed
with the Submillimeter Array (SMA) and the Caltech Submillimeter Observatory (CSO) covering pc-scale clouds
to mpc-scale cores, are analyzed in order to quantify the magnetic field importance. The magnetic
field misalignment $\delta$ -- the local angle between magnetic field and dust emission gradient --
is found to be a prime observable, revealing distinct distributions for sources where the magnetic
field is preferentially aligned with or perpendicular to the source minor axis. Source-averaged 
misalignment angles $\langle|\delta|\rangle$ fall into systematically different ranges, reflecting
the different source-magnetic field configurations. Possible bimodal $\langle|\delta|\rangle$-distributions
are found for the separate SMA and CSO samples. Combining both samples broadens the distribution with a 
wide maximum peak at small $\langle|\delta|\rangle$-values. Assuming the 50 sources to be representative, 
the prevailing source-magnetic field configuration is one that statistically prefers small magnetic field
misalignments $|\delta|$. When interpreting $|\delta|$ together with a magnetohydrodynamics force equation, 
as developed in the framework of the polarization-intensity gradient method, a sample-based
log-linear scaling fits
the magnetic field tension-to-gravity force ratio $\langle\Sigma_B\rangle$ versus $\langle|\delta|\rangle$ with 
$\langle\Sigma_B\rangle = 0.116 \cdot \exp(0.047\cdot \langle|\delta|\rangle)\pm 0.20$ (mean error), 
providing a way to estimate the relative importance of the magnetic field, only based on measurable
field misalignments $|\delta|$. The force ratio $\Sigma_B$ discriminates systems
that are collapsible on average 
($\langle \Sigma_B\rangle <1$) from other molecular clouds where the magnetic field still provides 
enough resistance against gravitational collapse ($\langle \Sigma_B\rangle >1$). The sample-wide 
trend shows a transition around $\langle|\delta|\rangle\approx 45^{\circ}$. Defining an effective 
gravitational force $\sim 1-\langle \Sigma_B\rangle$, the average magnetic-field-reduced star-formation
efficiency is at least a factor of two smaller than the free-fall efficiency. For about one forth of the 
sources the average efficiency drops to zero. The force ratio $\Sigma_B$ can further be linked to the normalized
mass-to-flux ratio, yielding an estimate for the latter one without the need of field strength 
measurements. Across the sample, a transition from magnetically supercritical to subcritcal is observed
with growing misalignment $\langle|\delta|\rangle$.

\end{abstract}
%
%
%
%

\section{Introduction}     \label{intro}

Magnetic fields and their role in star formation have long been studied in the literature
\citep[e.g.,][]{crutcher12, mckee07, shu06, shu04}. Polarization observations in the optical, 
infrared and submillimeter bands often routinely detect extended polarized emission, revealing
magnetic field morphologies across entire star-forming regions. While the higher-frequency
optical / near-infrared light is polarized 
as a result of scattering off dust grains \citep[e.g.,][]{santos14, pereyra08}, 
the submillimeter wavelengths 
directly show the emitted polarized light from dust. The grains are thought to be aligned with 
their shorter axes parallel to the magnetic field lines in most situations, thus emitting in a 
plane perpendicular to the field lines \citep{cudlip82, hildebrand84, hildebrand88, lazarian07, andersson12}.
Many submillimeter facilities have developed their polarization capabilities. Operating at a
wavelengths $\lambda \sim 350 \mu$m, Hertz on the 
Caltech Submillimeter Observatory (CSO) mapped magnetic fields in star-formation regions
with a resolution of about 20\arcsec. A summary catalog with 56 sources is given in \citet{dotson10}.
Its successor instrument, SHARP \citep{li08b}, is currently being operated on the CSO, achieving a resolution 
of about 10\arcsec. 
Dust continuum polarization was also observed with SCUPOL on the James Clerk Maxwell Telescope
(JCMT) at  $\lambda \sim 850 \mu$m with a binned resolution of about 9\arcsec, summarized in 
the legacy catalog with 83 sources by \citet{matthews09}. Higher-resolution maps are possible 
with interferometers. The  Berkeley-Illinois-Maryland Association (BIMA) array reached about 2 -4\arcsec
at 1.3 and 3.3~mm \citep{rao98,lai01,lai02}. More recently, both the Combined Array for Research 
in Millimeter-wave Astronomy (CARMA) and the Submillimeter Array (SMA) 
presented results from legacy-type surveys toward 38 star-forming 
regions and cores (\citet{hull14}, $\theta\sim 2.5$\arcsec, $\lambda \sim 1.3$mm) and 14
high-mass star-forming clumps (\citet{zhang14}, $\theta\sim 1-3$\arcsec, $\lambda \sim 870 \mu$m).

All these observations have in common that they often cover large enough areas (of polarized 
emission) such that connected patches of projected magnetic field orientations become visible. 
This is advantageous to investigate magnetic field morphologies. Consequently, several studies are
making use of this in combination with other tracers such as outflows, larger scale magnetic field 
orientations or theoretical models. The class 0 low-mass protostar L1157-mm is found to have 
an hourglass morphology (from CARMA observations) that aligns both with its outflow and also with 
the larger scale field detected by SHARP \citep{stephens13}. The magnetic field in the dense clump
IRAS 20126-4104 (SHARP, \citet{shinnaga12}) is both aligned with its bipolar outflow as well as 
the larger scale field from optical polarimetry data. A comparison with numerical simulations 
indicates a cloud rotation axis inclined by about $60^{\circ}$ with respect to the magnetic field. A 
positive correlation between mean field orientations and both pseudo-disk symmetry axes and 
bipolar outflow axes is derived from seven class 0 protostars in \citet{davidson11} and \citet{chapman13} based
on SHARP  polarization detections, sampling $\sim 10,000$~AU scales. On smaller scales, \citet{hull14}
find that a subset of their sources has consistent magnetic field orientations between the larger 
scale ($\sim 20$\arcsec, probed by single-dish submillieter telescopes) and the smaller scale 
($\sim 2.5$\arcsec) magnetic fields from CARMA. This is interpreted as the magnetic field playing 
a role in regulating infall down to smaller scales ($\sim 1000$~AU) in these sources. Bipolar
outflows (CO 2-1 transition) in their sample of low-mass protostars seem to be randomly oriented 
with respect to the smaller scale fields \citep{hull13, hull14}, although there is a trend that low-polarization 
sources have their outflows perpendicular to small-scale fields, indicating that these fields might
being wrapped by rotation \citep{hull14}. The SMA sample \citep{zhang14} indicates that 
magnetic fields at dense core scales are either aligned within $40^{\circ}$ of or perpendicular to 
the larger parsec-scale fields. Similar to the CARMA result, they also conclude that outflows 
(CO 3-2 transition) are randomly oriented with the core magnetic fields. 
Numerical simulations of collapsing cores, investigating scenarios to align / misalign rotation, 
outflows and magnetic fields are presented in e.g., \citet{li13, krasnapolsky12, machida11, ciardi10}.

In ideal situations, magnetic field theoretical models and simulations can be fairly 
closely matched to polarization observations \citep[e.g.,][]{goncalves08, frau11,kataoka12}.
A mathematical model to fit an observed hourglass field morphology is derived in 
\citet{ewertowski13}. Their analytic formulae -- assuming an axisymmetric purely poloidal field threading
a flattened star-forming core and connecting to a uniform background field -- are using the 
shape of the hourglass pattern to constrain the ratio of local-to-background magnetic field strength.
Provided an independent measurement of the background field strength, local field strength 
values in a region of interest can be calculated from their analytic approach.

Many observed star-forming sources are, however, irregular and complicated. It is, thus, important and 
desirable to develop new techniques that can directly and independently extract information 
from detected polarization. A major shortcoming of dust polarization is that only sky-projected
field morphologies (rotated by 90$^{\circ}$ from measured polarization orientations) are 
observed but no magnetic field strengths. Recent developments aiming at improving this situation 
include the polarization-angle dispersion function leading to an estimate of the turbulent-to-mean
magnetic field strength \citep{hildebrand09,houde09,koch10,houde11,poidevin13}, the study of 
turbulent velocity dispersion spectra of co-existing ions and molecules to constrain the ambipolar
diffusion scale and the plane-of-sky component field strength \citep{hezareh10, li08} and the 
polarization-intensity gradient method to derive the local total magnetic field strength
\citep{koch12a,koch12b}. In this work here, we focus on a previously introduced observable: the angle 
$\delta$ between the local magnetic field orientation and its underlying dust emission gradient 
orientation. Although not limited to dust polarization measurements, 
we focus on dust submillimeter observations across a large and diverse sample, with the 
goal to (1) illustrate characteristic features and differences  in $\delta$ based on a to date
largest sample of 50 sources, and (2) to provide
an interpretation for $\delta$ via the polarization-intensity gradient method, illustrating clear
trends in the importance of the magnetic field across the sample that are accessible via $\delta$.
This study is part of the program on the SMA \citep{ho04} to investigate the structure of the magnetic 
field from large to small scales.

The paper is organized as follows. Section \ref{section_sample} briefly describes selection 
criteria and properties of the SMA and the CSO sample. The main observable -- the angle 
$\delta$ -- its interpretation and the polarization - intensity gradient method are 
summarized in Section \ref{section_analysis}, based on earlier results in 
\citet{koch12a,koch12b,koch13}. Results for the angle
$\delta$ follow in Section \ref{section_result}, starting from overall-sample to group and 
source-averaged results. 
Section \ref{discussion} discusses interpretation and comparison with previous results.
Conclusion and summary are given in Section \ref{section_conclusion}.

\section{Samples}  \label{section_sample}

This work combines data from two different samples and telescopes (Table \ref{table_quantities}). 
Hertz on the single-dish
telescope CSO has published a catalog of polarization observations at a wavelength $\lambda \sim 350\mu$m
with a resolution $\theta\sim 20\arcsec$ \citep{dotson10}. The SMA, operating in various array configurations 
at $\lambda \sim 870\mu$m, is sampling smaller scales with resolutions of a few arcseconds down to 
sub-arcsecond scales.

\subsection{SMA}

The SMA\footnote{
The Submillimeter Array is a joint project between the Smithsonian Astrophysical Observatory and 
the Academia Sinica Institute of Astronomy and Astrophysics, and is funded by the Smithsonian
Institution and the Academia Sinica.
} has been pioneering high-resolution submillimeter polarization observations. Whereas the earliest
observations were focusing on the most promising and brightest targets 
\citep[e.g.,][]{girart06,girart09,rao09,tang09a,tang09b,chen12, tang12},
later receiver and correlator upgrades made it possible to also observe fainter sources. This has boosted
the sample of high-resolution low-and high-mass star-forming regions which currently totals about 30 sources.
A substantial number of these sources (14) was observed through a recent dedicated SMA polarization legacy 
program \citep{zhang14}, sampling star-forming clumps with densities larger than $10^5$~cm$^{-3}$
at scales around or smaller than 0.1~pc.
Detailed follow-up source-by-source analyses are being conducted, 
with first results published for DR21(OH) \citep{girart13}, G192.16-3.84 \citep{liu13}, G35.2-0.74 N \citep{qiu13}.
For our analysis here, we have applied selection criteria identical to the ones for the CSO sample: 
coherent polarization patches with at least 10 polarization segments, and a Stokes $I$
continuum detection strong enough to define a gradient.
Besides newly detected sources from the SMA polarization legacy program, the sample analyzed here
also comprises some earlier observed targets (Table \ref{table_quantities}).

\subsection{CSO}     \label{section_cso_sample}

The CSO/Hertz polarization catalog contains 56 regions, mostly Galactic sources and some galaxies 
\citep{dotson10}\footnote{
The publicly available reduced data can be found at 
{\rm http://iopscience.iop.org/0067-0049/186/2/406/fulltext/}.
}.
We chose a total of 27 sources from the CSO catalog.
Sources are selected based on their coverage of detected polarized emission, together with the condition
that the Stokes $I$ continuum emission need to be strong enough to define a gradient.
In order to have at least some connected patches, we request a minimum of 10 measured polarization --
intensity gradient pairs for our analysis. $|\delta|$-maps together with some basic 
statistical numbers for each of the CSO-selected sources are given in \citet{koch13}.

\section{Analysis Technique: Observable $\delta$ and Polarization - Intensity Gradient Method}
                                                                                                \label{section_analysis}

\begin{figure}
\begin{center}
\includegraphics[scale=1]{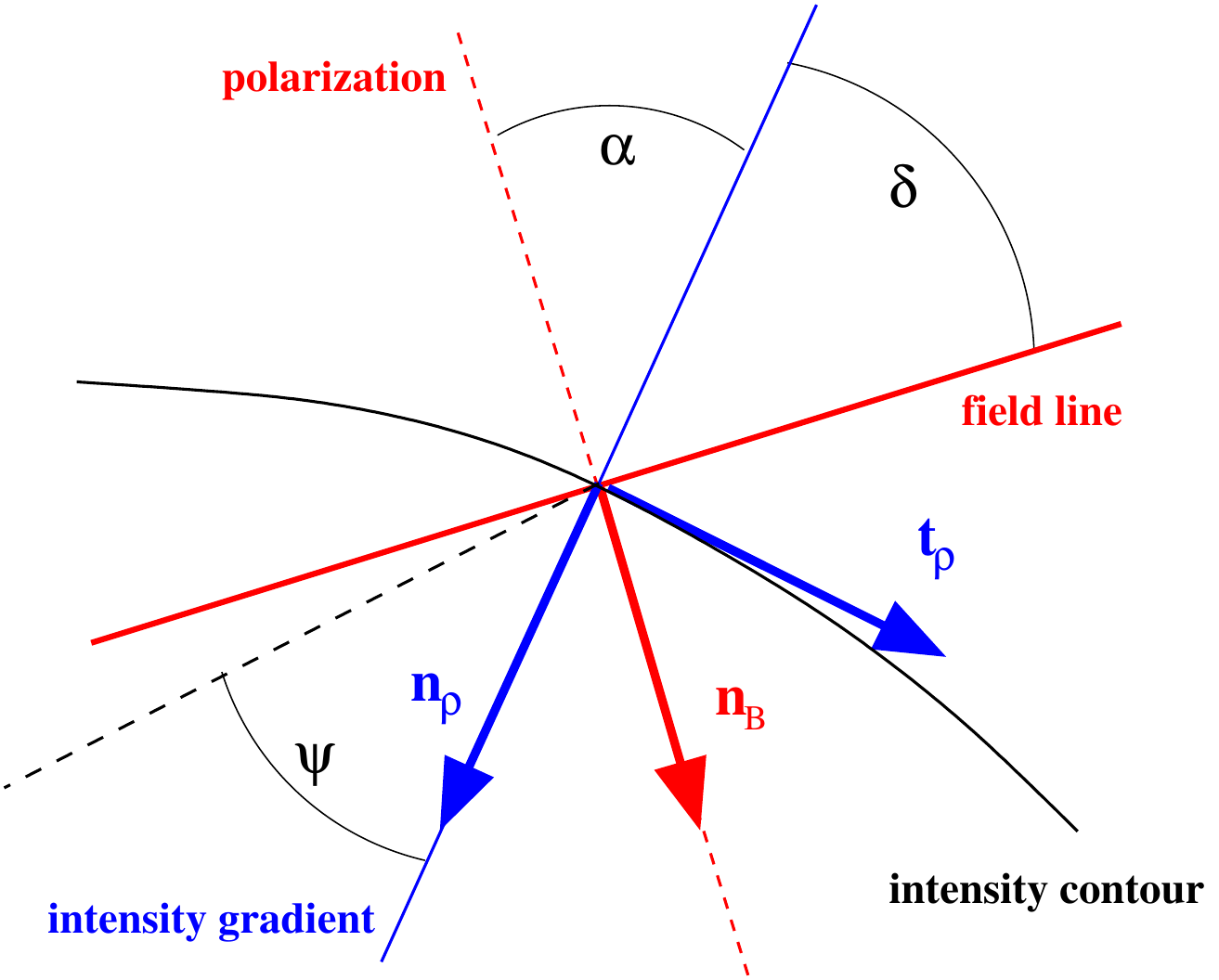}
 \caption{\label{schematic_simple} 
Schematic of the relevant angles to calculate the magnetic field significance
$\Sigma_B=\sin\psi/\sin\alpha$.
The angle $\delta$ is measured between magnetic field orientation (red solid line) and  
intensity gradient orientation (blue solid line), covering the range $0\le |\delta| \le 90^{\circ}$
or $-90^{\circ}\le \delta \le 90^{\circ}$ if additionally given a clockwise or 
counter-clockwise sense of rotation.
$\alpha$ is its complementary angle, i.e., $\alpha=\pi/2-|\delta|$.
The magnetic field tension force is directed normal to the field line 
along the unity vector $\mathbf{n}_B$, 
collinear to the originally detected 
dust polarization orientation (red dashed line) which is rotated by $90^{\circ}$ 
with respect to the magnetic field in the case of (sub-)millimeter 
dust polarization emission.  
The intensity gradient is normal to the 
emission intensity contour (black solid line) along the unity vector 
$\mathbf{n}_{\rho}\equiv \nabla\rho/|\nabla\rho|$, forming an angle $\psi$
with the (local) gravity and/or pressure gradient orientation
(black dashed line).
The unity vector tangential to the intensity contour, $\mathbf{t}_{\rho}$, forms an orthonormal 
system together with $\mathbf{n}_{\rho}$.
Figure reproduced from \citep{koch13}.
}
\end{center}
\end{figure}

The local angle $\delta$ is measured between a magnetic field orientation and a dust emission 
gradient (Figure \ref{schematic_simple}). In our study here, an observed dust polariztion orientation
which is then rotated by 90$^{\circ}$ reveals the magnetic field orientation. Nevertheless, any 
measurement showing the magnetic field morphology, as e.g., line polarization, Faraday rotation
or Zeeman splitting, can be used and what follows is generally applicable. $\delta$ is 
uanambiguously defined in the range 0 to $+90^{\circ}$ (as the smaller of the two angles 
complementary to $180^{\circ}$), or $-90^{\circ}$ to $+90^{\circ}$ when giving it a sense 
of orientation. $\delta$ is straighforward to derive from an observation. A simple inspection 
by eye can often distinguish between zones of small (prevailing field orientation rather orthogonal
to emission contours) and zones of large $\delta$ (prevailing field orientation rather parallel 
to contours). 

{\it What is the meaning of the angle $\delta$?} With the help of an ideal magnetohydrodynamics 
(MHD) force equation, by 
decomposing the magnetic field orientation in an orthonormal system with unity vectors parallel
and normal to an emission contour (Figure \ref{schematic_simple}), the angle $|\delta|$ with 
$\sin|\delta|  \in [0,1]$ quantifies the fraction of the available magnetic field tension force 
along a density gradient direction\footnote{
Observed emission -- generally a combined result of density and temperature -- does not 
directly correspond to the density as used in the MHD force equation. This is, nevertheless, 
not of a concern, because our approach involves the orientation of the gradient of 
emission or density. A constant temperature, steep or gentle temperature gradients that align
with a density gradient will all lead to emission gradient orientations that can be identified 
(in orientation) with the MHD density gradient orientation. Only a significant temperature gradient
that deviates a lot in direction from a density gradient could be problematic. 
}
\citep{koch13}. 
Thus, $|\delta|$ can be interpreted as a magnetic field alignment deviation or efficiency that
measures how the magnetic field is responsible for an observed density morphology along the 
gradient direction. Magnetic field lines closely aligned with density gradients, as e.g., seen in 
polar regions in collapsing systems, yield $|\delta| \approx 0$ or small, which indicates that 
the magnetic field tension force is not or only minimally slowing down infall in this region 
along the density gradient direction.  Contrary to that, in regions where the magnetic field is 
more bent, as e.g., in the inner part of an accretion plane where the field becomes close to 
orthogonal to a density gradient, $|\delta|$ is significantly larger than 0. Hence, the field tension
force can maximally inhibit collapse or any motion along the density gradient direction. 
Therefore, $|\delta|$-maps contain information on the role of the magnetic field \citep{koch13}.
Changes in $|\delta|$ across a star-formation region can reveal zones of major or minor 
magnetic field influence. Additionally, many sources reveal bimodal distributions in $\delta$ when
giving it a sense of orientation, i.e., when differentiating between a field line rotated clockwise
or counter-clockwise with respect to an emission gradient. This can be interpreted as signs 
of a collapse (or the beginning of a collapse) which will start to bend initially straight magnetic 
field lines along the accretion direction that subsequently leads to $+/-\delta$ above and 
below the accretion plane \citep{koch13}. 

We note that $\delta$ and its characteristic features, found across a sample of about 30 sources
in \citet{koch13}, are based entirely on observations only. In order to further quantitatively assess
the magnetic field, some of the observational results in the following sections will be further 
analyzed with the polarization-intensity gradient method \citep{koch12a}. At every position of 
detected polarization, this method leads to a local magnetic field strength and a ratio between the 
local field tension force and gravity. The background of this method is the ideal MHD force 
equation, making use of the misalignment angle $|\delta|$. In a molecular cloud, the interaction 
of all the forces (magnetic field, gravity, pressure gradients) leads to the observed magnetic 
field and dust morphologies. Hence, these forces leave a geometrical imprint, like the 
misalignment angle $|\delta|$, that can be identified with the various terms in the MHD force
equation. In particular, the direction of the field tension force is orthogonal to the detected 
field orientation (i.e., parallel to the detected dust polarization orientation), a field curvature
can be derived from adjacent field segments, the direction of the local gravitational pull can 
be calculated by summing up all the surrounding mass. With this, a force triangle can be 
constructed at every location of detected polarization. This triangle can be closed and solved
for the local field strength under the approximation that the orientation of the emission gradient
is a measure for the inertial term in the MHD force equation. By taking additionally into account
velocity bulk directions, this approximation was further studied in \citet{koch13} and confirmed
to be robust for field strength and force ratio estimates. The resulting local field strength is 
given by \citep{koch12a}:
\begin{equation}
B = \sqrt{\frac{\sin\psi}{\sin\left(\frac{\pi}{2}-|\delta|\right)} (\nabla P + \rho \nabla \phi) 4\pi\, R},
\end{equation}
where $\rho \nabla \phi$ and $\nabla P$ are the gravitational
pull and pressure gradient, and 
$\psi$ is the difference in orientation between the local gravitational pull and/or pressure
gradient and the emission
gradient (Figure \ref{schematic_simple}).
$\rho$ is the density and $R$ is the magnetic field radius. It is 
important to remark that this method provides a total magnetic field strength. With the field 
tension force $F_B = B^2/(4\pi R)$, the above equation can be rewritten as:
\begin{equation}
\Sigma_B \equiv \frac{\sin\psi}{\sin\left(\frac{\pi}{2}-|\delta|\right)} = \frac{F_B}{|F_G+F_P|},
\end{equation}
where $F_G$ and $F_P$ are gravitational pull and any possible additional pressure gradient.
$\Sigma_B$ quantifies how significant the local field tension force is, as compared to the 
other forces. Interestingly, this ratio is free of any molecular cloud model assumptions and 
only relies on two measurable angles that reflect the geometrical imprint of magnetic field 
and other forces. 
Assuming pressure gradients to be typically small\footnote{
Thermal pressure is balancing gravity following $GM/R_c \sim k_B T/m$ where $M$, $R_c$
and $T$ are the molecular cloud mass, radius and temperature. $m$ is related to the cloud
gas composition and proportional to the hydrogen mass, $m\sim m_H$, with a numerical
factor of order unity. For pure thermal support, this leads to 
$R_c\approx 0.1 (M/M_{\odot})\, (10 K/T)\,$pc. The sources in our samples comprise hundreds
up to thousands of solar masses. This leads to thermally supported length scales of $\sim$10~pc
or larger. Our largest observed scales in the CSO sample are about 1~pc, with most of the sources
being observed on scales of $\sim 10 - 100$~mpc. Thus, we are very likely sampling sufficiently
small scales where thermal pressure is small and negligible compared to gravity. This also still holds
if the molecular cloud temperature is higher with $T\sim$20-30~K.
Yet an additional pressure gradient can result from large-scale flows. 
Compressive flows are suggested to contribute to and shape observed filamentary networks
with coherent structures on pc- or larger scales \citep[e.g.,][]{peretto12, andre13}.
Significant pressure gradients might be present here, in particular for supersonic flows
and turbulence. Omitting such gradients can lead to uncertainties in our analysis.
This shortcoming can be overcome if additional large-scale velocity information is
available overlapping with polarization data. The sample analyzed here, however, is again probing
mostly smaller scales which likely eliminates a possibly significant bias from large-scale converging flows.
Finally, pressure from turbulence with velocity $v_T$ scales as $v_T^2 \sim k_BT/m$.
With typical velocity dispersions of  $\sim1-2$~km/s in our sample, gravity is also 
dominating over turbulent pressure on scales of $\sim1$~pc and smaller.}
compared to gravity in our samples, $\Sigma_B$
provides a criterion whether local collapse is possible ($\Sigma_B < 1$) or whether the field 
tension is large enough to inhibit local collapse ($\Sigma_B > 1$). $\Sigma_B$-maps of 
molecular clouds systematically show zones of smaller- or larger-than-one $\Sigma_B$-values.
The same transition can often also be observed in $|\delta|$-maps which makes them an 
easy tracer for changes in the field significance. 
$|\delta|$- and $\Sigma_B$-maps for each of the CSO-selected sources (Section \ref{section_cso_sample}
and Table \ref{table_quantities}) are given in \citet{koch13}. 
Finally, both $\Sigma_B$
and $B$ are little or not at all affected by projection effects. This is possible because the 
projection correction enters as a ratio of two inclination angles (for $\psi$ and $|\delta|$). 
For identical inclination angles for magnetic field line and emission gradient, the projection 
effect cancels out. The projection effect is negligible or less than 30\% as long as one 
inclination angle is within 45$^{\circ}$ and the other one is not more than 50$^{\circ}$
different from the first one \citep{koch12a}.

\section{Results}  \label{section_result}

This section starts by presenting SMA-CSO-combined overall results (Section \ref{result_sample}).
This is followed by a refined analysis where sources from the combined samples are categorized
based on proposed criteria (Section \ref{result_category}). Finally, Section \ref{result_individual}
looks into averaged source-by-source trends and compares them to the proposed categorization from 
Section \ref{result_category}. 
Results for both all independent $\delta$, $\psi$ and $\Sigma_B$, and also 
for their source-averaged quantities $\langle\cdots\rangle$ are shown.

\subsection{Overall Sample} \label{result_sample}

\begin{figure}
\begin{center}
\includegraphics[scale=0.7]{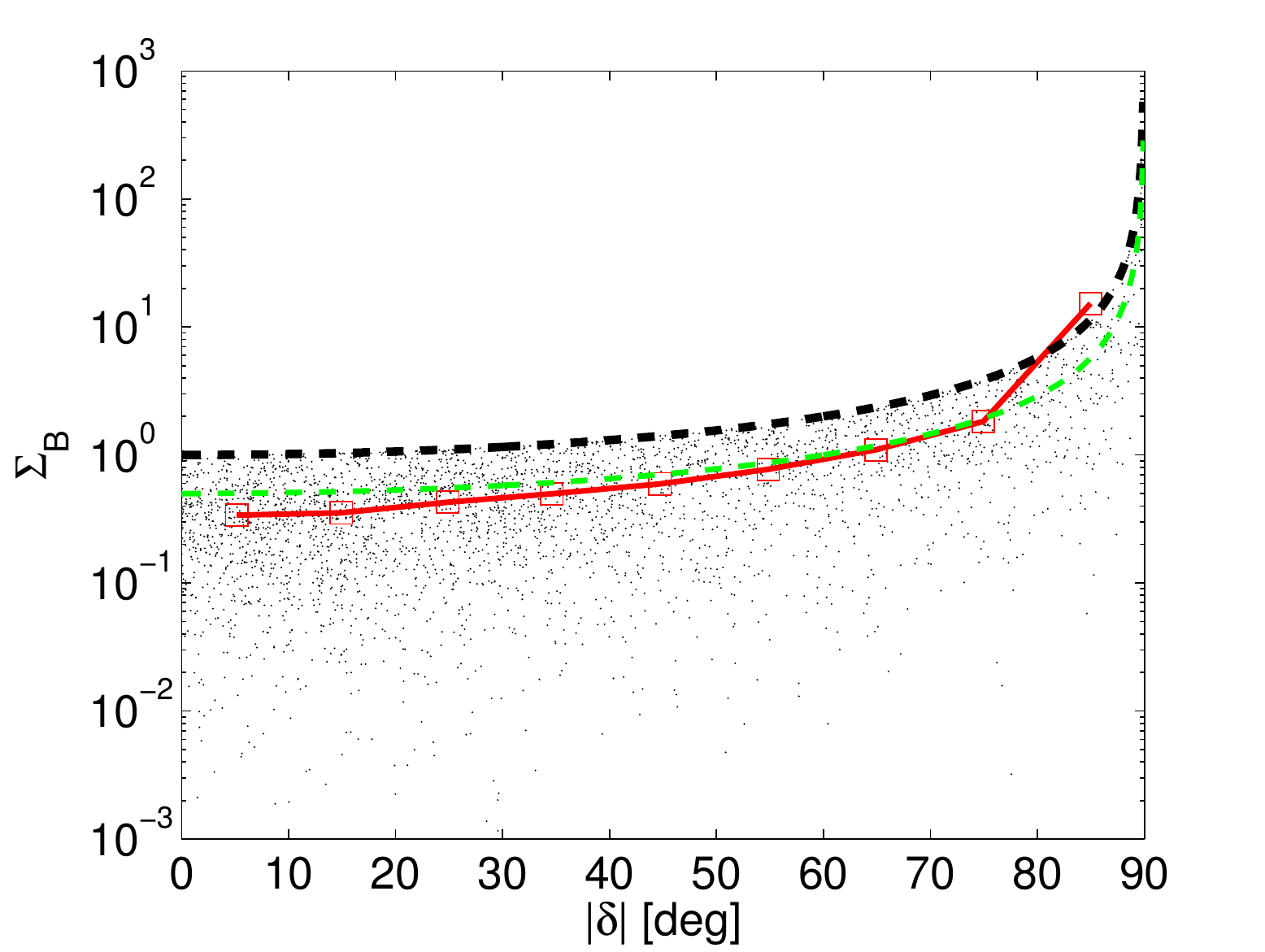}
 \caption{\label{figure_sample_scatter} 
$\Sigma_B$ vs $|\delta|$ for all sources from the combined SMA and CSO samples
(close to 4,000 independent ($|\delta|,\Sigma_B$)-pairs).
Bin-averaged values are shown with red squares and are connected with straight lines.
Values are binned in 10$^{\circ}$ intervals. The black dashed line shows the upper
envelope $1/\sin\alpha \ge 1$. The averaged value in the last bin around $85^{\circ}$
falls beyond the upper envelope because of the exponential growth in $\Sigma_B$ 
for $|\delta|\rightarrow 90^{\circ}$.
The dashed green line is $1/2 \cdot 1/\sin\alpha$.
}
\end{center}
\end{figure}

We start by looking for general overall trends, based on the combined full SMA-CSO sample.
Here, we intentionally do not distinguish between SMA and CSO sources or any further source 
properties because we are aiming at characterizing the magnetic field in any star-forming region.
To this purpose, we show for all sources with all their detected polarization 
orientations, their angles $|\delta|$ with their corresponding $\Sigma_B$-values 
(Figure \ref{figure_sample_scatter}). For the total of 50 sources, this amounts to close to 
4,000 independent ($|\delta|, \Sigma_B$)-pairs. This scatter plot clearly traces and follows an upper envelope
which is determined by the maximum possible $\Sigma_B$-values which are $max\, \Sigma_B=1/\sin\alpha \ge 1$
for $\sin\psi \equiv 1$. 
The combined sample fills the entire possible range in $|\delta|$ with a distribution displaying
a broad peak around smaller $|\delta|$-values with a slow roll-off and an extended tail towards
larger values (lower panel in Figure \ref{source_psi_delta}).
Binning $|\delta|$ into intervals of 10$^{\circ}$ shows uniform $|\delta|$-distributions with gradually
fewer data points in subsequent bins towards larger $|\delta|$-values.
The distributions of $\Sigma_B$ show broad peaks around $\Sigma_B$-values initially smaller 
than 1. These peaks broaden in subsequent $|\delta|$ bins and shift eventually toward $\Sigma_B$-values
larger than 1.
At this point here, no distinction is made between individual sources. As 
we will discuss in the Sections \ref{result_category} and \ref{result_individual}, 
distinct source types preferentially fall into distinct zones in the scatter plot in 
Figure \ref{figure_sample_scatter}. 
Filling the entire possible
ranges in $|\delta|$ and $\Sigma_B$ possibly indicates that we might have a fairly complete and unbiased sample
representing all possible source types.
We stress that it is not automatically obvious from 
the beginning that the scatter plot (Figure \ref{figure_sample_scatter}) needs to show a complete
(though not uniform) coverage. This is not obvious because the force ratio $\Sigma_B=\sin\psi/\sin\alpha$ allows 
for a vast range of $\Sigma_B$-values within $[0,\infty)$. In particular, values could be exclusively always smaller 
than 1, or they could be always $\ge 1$. Nevertheless, values 
cover a range across the dividing line  $\Sigma_B=1$.
While a single angle $|\delta|$ can be affected by projection effects due to its unknown
inclination angle, for a large number of more than 1,000  independent data points
projection effects are likely averaging out.
The binned uniform $|\delta|$-distributions together with the gradually broadening 
$\Sigma_B$-distributions
suggest that, generally on average, larger $|\delta|$-values
point toward larger $\Sigma_B$-values.  And indeed, $|\delta|$-values averaged for all the binned
intervals confirm this. Furthermore, the binned values follow approximately the curve 
$\sim 1/2\cdot 1/\sin\alpha$ (Figure \ref{figure_sample_scatter}), which is simply the mean
between 0 and  $max \,\Sigma_B=1/\sin\alpha$.

In conclusion, for a large sample without further distinguishing any individual sources, 
we find a sample-wide trend where growing $|\delta|$-values indicate growing $\Sigma_B$-values
following $\Sigma_B\approx 1/2\cdot 1/\sin(\pi/2-|\delta|)$. On average, gravity is balanced by
the magnetic field tension force, i.e., $\Sigma_B \equiv 1$, for $|\delta|=60^{\circ}$. 
Individual sources or groups of sources can still reveal different properties 
(Section \ref{result_category} and \ref{result_individual}).

\subsection{Source Categorization} \label{result_category}

\begin{figure}
\begin{center}
\includegraphics[scale=0.42, angle=90]{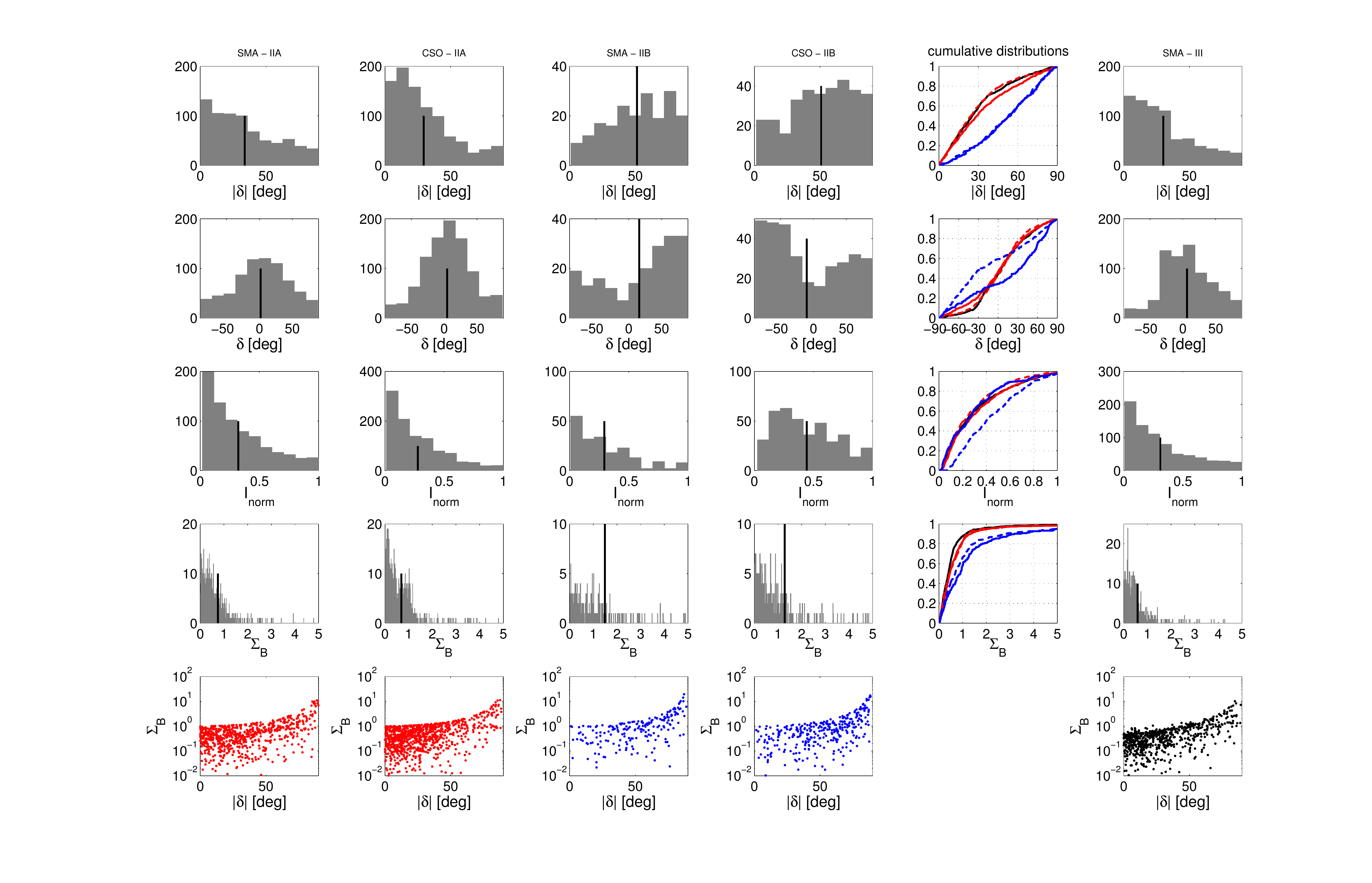}
 \caption{\label{hist_cat} \tiny
Histograms for categorized source groups: 
the columns one and two are for SMA-IIA and CSO-IIA, columns three and four
are for SMA-IIB and CSO-IIB and column six is for SMA-III. 
Distributions for observed absolute $|\delta|$, relative $\delta$ 
and $I_{norm}$ are shown in the top, second and third row.
Clear differences are apparent.
$\Sigma_B$-histograms, derived from the 
polarization-intensity gradient method, are displayed in the forth row. Solid vertical lines 
mark mean values. The bottom row shows scatter plots of $\Sigma_B$ vs $|\delta|$, illustrating 
how the different categories fill in different areas in this parameter space.
Column five displays cumulative distributions for all the histograms for 
$|\delta|$, $\delta$, $I_{norm}$ and $\Sigma_B$
for IIA (red), IIB (blue) and III (black) where solid lines are used for the SMA
and dashed lines for the CSO distributions.
}
\end{center}
\end{figure}

\citet{koch13} proposed a scenario with an evolutionary magnetic field role which 
leaves its imprints in the angle $|\delta|$. 
Two distinct intermediate phases are identified (Table \ref{table_source_type}):
sources of type IIA (with their magnetic field roughly perpendicular to the cloud major axis), 
and sources of type IIB (with their magnetic field roughly parallel to the cloud major axis). 
This categorization was motivated by the very clear examples of Mon R2 (IIA) and M$+0.25+0.01$ (IIB) 
from the CSO catalog \citep{dotson10}. Additional sources from the CSO sample were then 
tentatively categorized in Table 2 in \citet{koch13}. For completeness, an extended version of 
this table including the new SMA sources is reproduced here (Table \ref{table_quantities}).
Since we are also interested in inter-sample differences (between the SMA and the CSO) we
show results here separately for the two instruments. 
The top row in Figure \ref{hist_cat} shows $|\delta|$-histograms for all the sources that are
(clearly) identified as IIA and IIB. This includes 12 IIA and 7 IIB sources from the CSO sample
and 9 IIA and 4 IIB sources from the SMA sample. 
The remaining sources are discarded because of 
their less complete and/or less clear
polarization patterns\footnote{
Table \ref{table_quantities} still lists an attempted tentative categorization for all the 
sources in both samples.
}. 
Individual source-based histograms in Appendix B in \citet{koch13} for the CSO sample
were already hinting
differences in distributions (which was an additional motivation for the classification scheme).
The group-based histograms now reveal a very clear difference with a falling distribution
with $\langle |\delta| \rangle_{IIA} = 34^{\circ}$ (SMA) and 30$^{\circ}$ (CSO), 
and an opposite trend with a growing
distribution with $\langle |\delta| \rangle_{IIB} = 51^{\circ}$ (SMA) and 51$^{\circ}$ (CSO).
As an obvious consequence, 
a Kolmogorov-Smirnov (KS) test rejects the null hypothesis (i.e., probability $p$ that the 
two samples IIA and IIB originate from an identical distribution) with $p\sim 10^{-15}$ (SMA)
and $p\sim 10^{-37}$ (CSO), 
where the maximum deviations ($k$-value) between their cumulative distributions are $k=0.34$ (SMA)
and $k=0.42$ (CSO). 
An inter-sample KS test, i.e., IIA$_{SMA}$ versus IIA$_{CSO}$ and IIB$_{SMA}$ versus IIB$_{CSO}$,
gives $p\sim 10^{-4}$ and $k=0.10$ for IIA and $p=0.94$ and $k=0.05$ for IIB.
We can, thus, conclude that equal-type distributions from two different instruments show a 
much closer resemblance than distributions from different types from the same instrument,
with IIB-type sources even showing  a high probability to reflect identical distributions. This 
is intriguing because this indicates that the sampling of different scales and densities by the SMA
and the CSO is less relevant in the context here. Independent of that, both instruments seem to reveal the same 
different types of morphologies. This can also be seen from the cumulative distributions in 
column 5 in Figure \ref{hist_cat}. The last column shows the results for 5 type-III\footnote{
Evolving from the more elongated structures of IIA- and IIB-type sources, the later type-III sources
are less elongated, but more symmetrized as gravity starts to pull in the magnetic field lines \citep{koch13}.
} sources from 
the SMA sample. No type-III sources are found in the CSO sample. The $|\delta|$-
distribution shows an even more pronounced peak towards small values with 
$\langle|\delta|\rangle\approx 30^{\circ}$.

Besides the angle $|\delta|$, the Stokes $I$ integrated dust emission is another prime observable
that reflects mass and morphology in the star-forming regions. Since every source has a different
absolute flux emission, a peak-normalized emission, $I_{norm}\in [0,1]$, is used in order to compare 
emission distributions across samples. The third row in Figure \ref{hist_cat} displays 
histograms of $I_{norm}$ for IIA and IIB in the SMA and CSO samples.
For these distributions, the normalized Stokes $I$
emission only at the positions of detected polarized emission (where $|\delta|$ is identified)
is used. As it is evident from the two CSO panels, the $I_{norm}$-histograms for IIA and IIB are 
very different among the CSO sample:
the IIA-group shows an exponential-like fall off toward the stronger emission 
end, whereas type-IIB sources reveal a broader distribution in $I_{norm}$ with a more gentle
decrease with brighter emission. A KS test again clearly discriminates between the two 
distributions ($p\sim 10^{-22}$, $k=0.29$). The mean normalized emissions are 
$\langle I_{norm}\rangle_{IIA} = 0.28 $ and $\langle I_{norm}\rangle_{IIB} = 0.44 $. The 
observed Stokes $I$ emission is an integrated and projected quantity. Nevertheless, the clear
difference in distributions between IIA and IIB is likely pointing to some underlying differences
in morphology; e.g., IIB CSO sources show more abundant brighter areas than IIA sources, which could
be a consequence of flatter or steeper density profiles.
All SMA types (IIA, IIB, III) show exponential-like fall offs 
where all average values are around 0.30.
All the SMA distributions are fairly similar
to the CSO-IIA distribution, leaving only the CSO-IIB distribution as clearly distinct 
(blue dashed line in the cumulative distribution plot in the third row. 

With the clear segregation in $|\delta|$, the magnetic field seems a plausible
candidate to explain these. Therefore, 
we proceed to analyze and interpret the above observational findings with the polarization--
intensity gradient method \citep{koch12a}. Histograms for the field significance $\Sigma_B$ are
shown in the fourth row in Figure \ref{hist_cat}. $\Sigma_B$-maps were previously published in 
Appendix B in \citet{koch13} for the CSO sources.
For the histograms in Figure \ref{hist_cat}, a sample-based (IIA, IIB)
$3\sigma$-clip is applied to the $\Sigma_B$-values in order to remove outliers\footnote{
This does in no way change the shape of the distributions, but it can change average values. 
A single excessively large $\Sigma_B$-value of the order of $\sim 100$ can significantly boost
$\langle\Sigma_B\rangle$ because typical $\Sigma_B$-values are of the order $\sim 1$. We note
that outliers are always a consequence of the exponentially growing $\Sigma_B \sim 1/\sin\alpha$
for small angles $\alpha (=\pi/2-|\delta|)$.
}
. Similarly, values below $\Sigma_B\equiv 0.01$ (i.e., 1\% field significance) are removed. Both 
groups of IIA sources show more uniform distributions up to about $\Sigma_B \sim 1$, followed
by a rather sharp drop. Type-IIB groups reveal a more exponential-like tail with a 
significant contribution from the range $\Sigma_B \sim 1 - 2$. A KS test identifies the two types of
distributions as clearly different (CSO IIA versus IIB: $p\sim 10^{-8}$, $k=0.19$;
SMA IIA versus IIB: $p\sim 10^{-11}$, $k= 0.29$).
Average values 
differ with $\langle \Sigma_B \rangle_{IIA} = 0.74$ (SMA) and $\langle \Sigma_B \rangle_{IIB} = 1.49$ (SMA)
and 
$\langle \Sigma_B \rangle_{IIA} = 0.69$ (CSO) and $\langle \Sigma_B \rangle_{IIB} = 1.29$ (CSO). 
The SMA III sources show an even more pronounced peak at smaller $\Sigma_B$-values with 
$\langle\Sigma_B \rangle=0.59$. The systematic difference between IIA, IIB and III is also 
clearly seen in the cumulative distributions with an almost complete absence of values $\Sigma_B > 1$
for III.
Thus, the observed differences in $|\delta|$ between the samples IIA and IIB seem to point to 
a different role of the magnetic field, where $\Sigma_B$ provides a way to quantitatively 
discriminate between the samples, identifying IIA ($ \langle \Sigma_B\rangle < 1$) and IIB
($\langle\Sigma_B\rangle > 1$) as potentially collapsing and non-collapsing systems on average. The bottom 
row in Figure \ref{hist_cat} illustrates this result with $\Sigma_B$-versus-$|\delta|$ plots: 
IIA sources scatter more densely around smaller $|\delta|$-values, while IIB sources scatter
around larger $|\delta|$-values. Type-III sources are even more concentrated around smaller
$|\delta|$-values (Table \ref{table_source_type}).

\subsection{Source-Averaged Analysis} \label{result_individual}

\begin{figure}
\begin{center}
\includegraphics[scale=0.55]{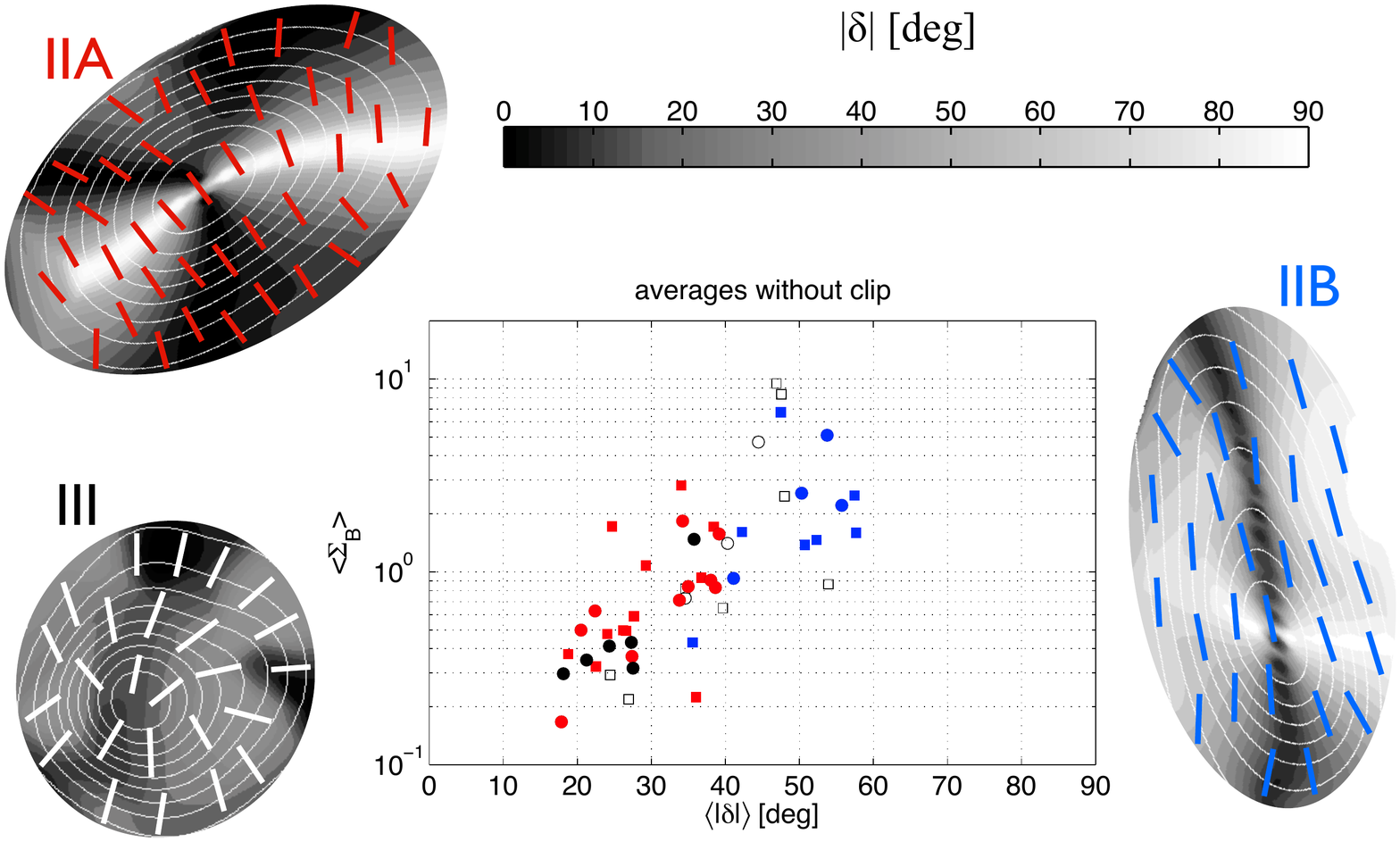}
 \caption{\label{figure_schematic} 
Source-averaged $\langle \Sigma_B\rangle$-values vs source-averaged $\langle |\delta|\rangle$-values
for the combined SMA and CSO samples. Red and blue filled symbols show
the clearly identified IIA- and IIB-type sources from both the SMA (circles) 
and CSO (squares) sample (Table \ref{table_quantities}).
Empty symbols correspond to sources with less clear features.
Black filled circles are the clearly identified SMA-III sources. 
Averages are calculated including outliers.
Schematic magnetic field structures are illustrated for IIA- (red segments), 
IIB- (blue segments) and III-type sources (white segments).  White contours display
the dust continuum emission.  The black-to-white color grading on the top indicates the local $|\delta|$.
Clear opposite trends are apparent for type-IIA and -IIB sources.
All 3 schematic types are generated by over-gridding and interpolating really observed
sources (IIA, IIB: G240 and G35.2 from \citet{zhang14} and \citet{qiu13}; III: W51 e2 from \citet{tang09b}).
The systematic local changes in $|\delta|$-maps reflect changes in $\Sigma_B$-maps \citep{koch13} which 
explains the observed correlation between $\langle \Sigma_B\rangle$ and $\langle |\delta|\rangle$ across the two samples.
Properties for the various source types are summarized in Table \ref{table_source_type}.
$\Sigma_B$-maps are shown in Figure \ref{figure_schematic_Sigma_B}.
}
\end{center}
\end{figure}

\begin{figure}
\begin{center}
\includegraphics[scale=0.6]{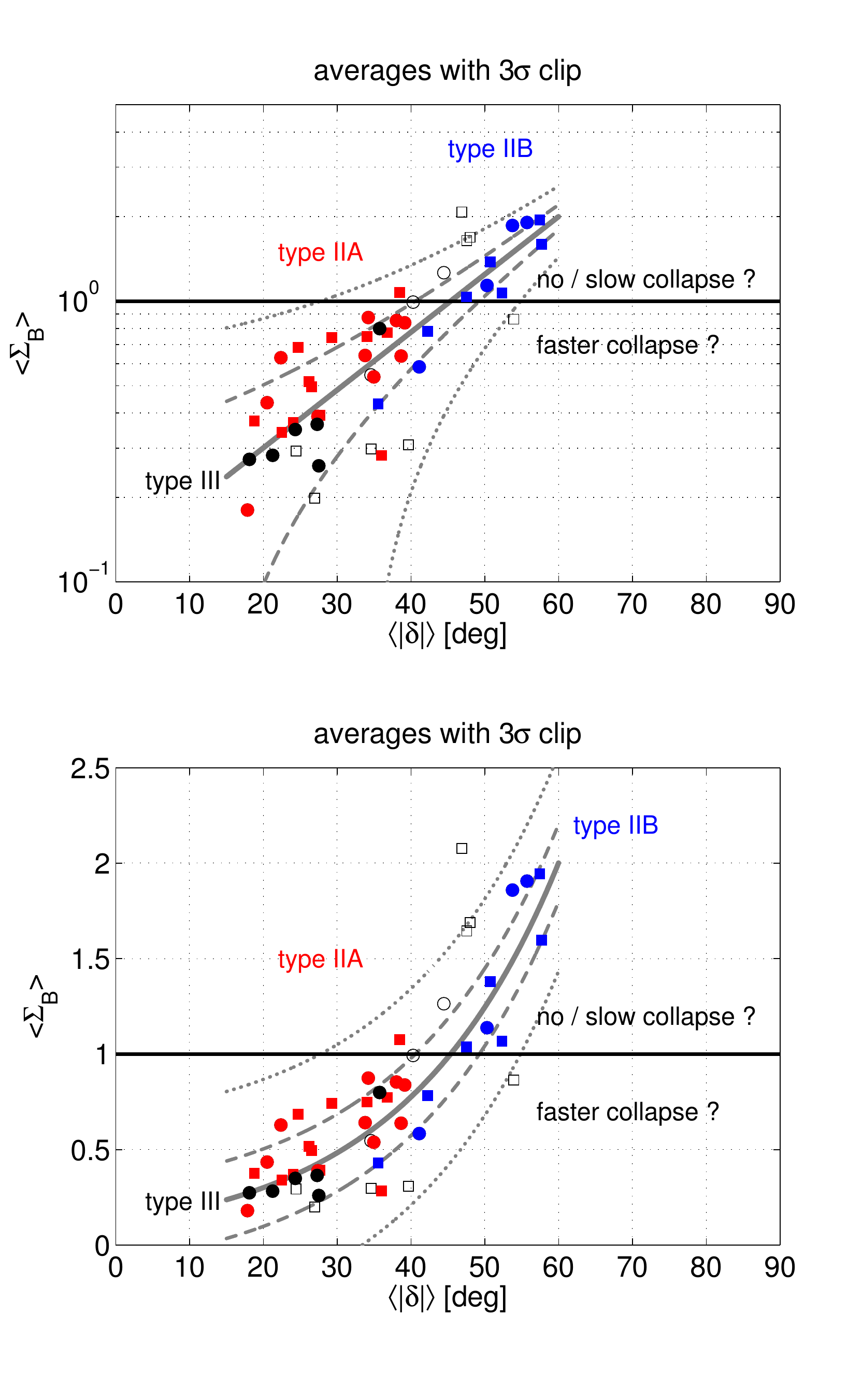}
 \caption{\label{source_Sigma_B_delta} \footnotesize
Source-averaged $\langle \Sigma_B\rangle$-values vs. source-averaged 
     $\langle |\delta|\rangle$-values for the combined SMA and CSO samples. Different from Figure   
     \ref{figure_schematic}, outliers in $\Sigma_B$ are removed for each source
     individually with a $3\sigma$ clip. Symbols and colors are identical to the ones in Figure  
     \ref{figure_schematic}. The horizontal line marks the force equilibrium 
     $\langle\Sigma_B\rangle \equiv 1$, below and above which gravity, {\it on average}, 
     is or is not overwhelming the magnetic field.
     $\langle\Sigma_B\rangle $ possibly discriminates between systems with, {\it on average},
     no / slow collapse and faster collapse.
     In either case, {\it local collapse} and  {\it local star formation} 
     can be possible or hindered as $\Sigma_B$-maps show
     distinct zones with $\Sigma_B$ larger or smaller than one 
     (Figure \ref{figure_schematic_Sigma_B} and \citet{koch13}).
     Clearly identified IIA-type (red) and IIB-type (blue) sources appear to fall onto the left and right
     side of the transition around $\langle |\delta|\rangle \sim 40^{\circ}$.
     The solid gray line is the log-linear best fit 
     $\langle\Sigma_B\rangle = A\cdot \exp(B\cdot \langle|\delta|\rangle)$ with 
     $A=0.116$ and $B=0.047$. The dashed and dotted gray lines indicate mean error
    ($\pm \,0.20$) and $3\sigma$ ($\pm \, 3\cdot 0.19$) bounds.
     The lower panel is identical to the upper panel but in linear-linear scale.
}
\end{center}
\end{figure}

From the category-based properties in Section \ref{result_category}, 
we now switch our focus to the analysis of individual
sources. Figure \ref{figure_schematic} displays source-averaged $\langle\Sigma_B\rangle$-values
as a function of the source-averaged $\langle|\delta|\rangle$-values. A $3\sigma$-clip is applied 
to each source separately in Figure \ref{source_Sigma_B_delta} 
in order to calculate $\langle\Sigma_B\rangle$. As a result, IIA sources
almost exclusively are found to have $\langle\Sigma_B\rangle < 1$, with $\langle |\delta|\rangle < 40^{\circ}$.
The majority of the IIB sources shows $\langle\Sigma_B\rangle > 1$, with $\langle |\delta|\rangle$ 
in the range between $\sim  40^{\circ}$ to $\sim  60^{\circ}$. This finding is not entirely
surprising after the clearly different distributions in Figure \ref{hist_cat} for IIA and IIB.
Nevertheless, it is not necessarily automatically clear that ensemble-based statistical properties
of IIA ($\langle\Sigma_B\rangle_{IIA} < 1$) and IIB ($\langle\Sigma_B\rangle_{IIB} > 1$) also apply
to each individual source in each category. It is, thus, rather striking that there seems to be a 
clear separation between IIA and IIB sources which identifies almost every source with $\langle \Sigma_B\rangle <1$
as IIA and $\langle \Sigma_B\rangle >1$ as IIB. Additionally, it is neither obvious from 
Figure \ref{hist_cat} that average $\langle|\delta|\rangle$-values fall strictly on one or the 
other side of $\langle|\delta|\rangle \approx  40^{\circ}$. 

We have to note that we have limited
the above description to the cases that were visually -- based on the magnetic field versus source 
major axis -- clearly categorized as either IIA or IIB (red and blue filled symbols in 
Figure \ref{figure_schematic} and \ref{source_Sigma_B_delta}).
What about sources that show less obvious patterns with, e.g., incomplete polarization coverages, 
signs of fragmentation or distorted and less organized field morphologies? 
The Figures \ref{figure_schematic} and \ref{source_Sigma_B_delta} show
$\langle\Sigma_B\rangle$ for all the SMA and CSO sources
listed in Table \ref{table_quantities}, independent of instrument, attempted categorization or
physical scale. Ignoring any classification scheme (Table \ref{table_source_type}),
a clear correlation of $\langle\Sigma_B\rangle$
with $\langle|\delta|\rangle$ is seen as well. We stress that no IIA / IIB-categorization is 
needed to arrive at this result. Our originally proposed distinction between (clear) IIA- and IIB-type
sources is based on a crude visual inspection only. Nevertheless, it seems that this simple by-eye
assessment is able to fairly reliably distinguish between the two regimes, $\langle\Sigma_B\rangle < 1$
or $\langle\Sigma_B\rangle > 1$ in Figure \ref{figure_schematic} and \ref{source_Sigma_B_delta}. In return, this finding
indicates: (1) Generally, if magnetic field lines tend to be perpendicular to a clump / core major
axis, gravity is dominating on average and overwhelming the magnetic field tension force ($\langle\Sigma_B\rangle < 1$).
(2) Generally, if the magnetic field lines tend to be aligned with 
a clump / core major axis, the magnetic field tension force is on average dominating over gravity.
This by-eye inspection is more precisely and quantitatively assessed with the 
angle $|\delta|$ which measures the alignment between magnetic field lines and clump / core
emission gradients. From Figure \ref{source_Sigma_B_delta}, the transition between possibly collapsible  
and non-collapsible systems, on average, occurs at $\langle|\delta|\rangle \approx 40^{\circ}$. Furthermore, 
it is interesting to note that sources that can not be unambiguously categorized 
(empty symbols in Figure \ref{figure_schematic} and \ref{source_Sigma_B_delta}) also follow the same
trend and fit in together with the categorized sources. Additionally, we mark sources that we 
categorize -- again simply by eye -- as hourglass-like with a pinched field morphology
(type III in our evolutionary scenario) with black filled symbols in Figure \ref{figure_schematic} 
and \ref{source_Sigma_B_delta}.
These sources presumably hint collapsing systems in an already more evolved stage (as compared to 
IIA and IIB). Consistent with that expectation, all these sources show a clearly dominating
gravity force with $\langle\Sigma_B\rangle$ around 30\% only, together with 
$\langle|\delta|\rangle \approx 20-30^{\circ}$ which is among the smallest values across the
entire sample. Fitting a straight line to all the data (SMA, CSO; type IIA, IIB, III) gives
$\langle\Sigma_B\rangle = 0.116 \cdot \exp(0.047\cdot \langle|\delta|\rangle)$ (Section \ref{discussion_fit}).

\begin{figure}
\begin{center}
\includegraphics[scale=0.55]{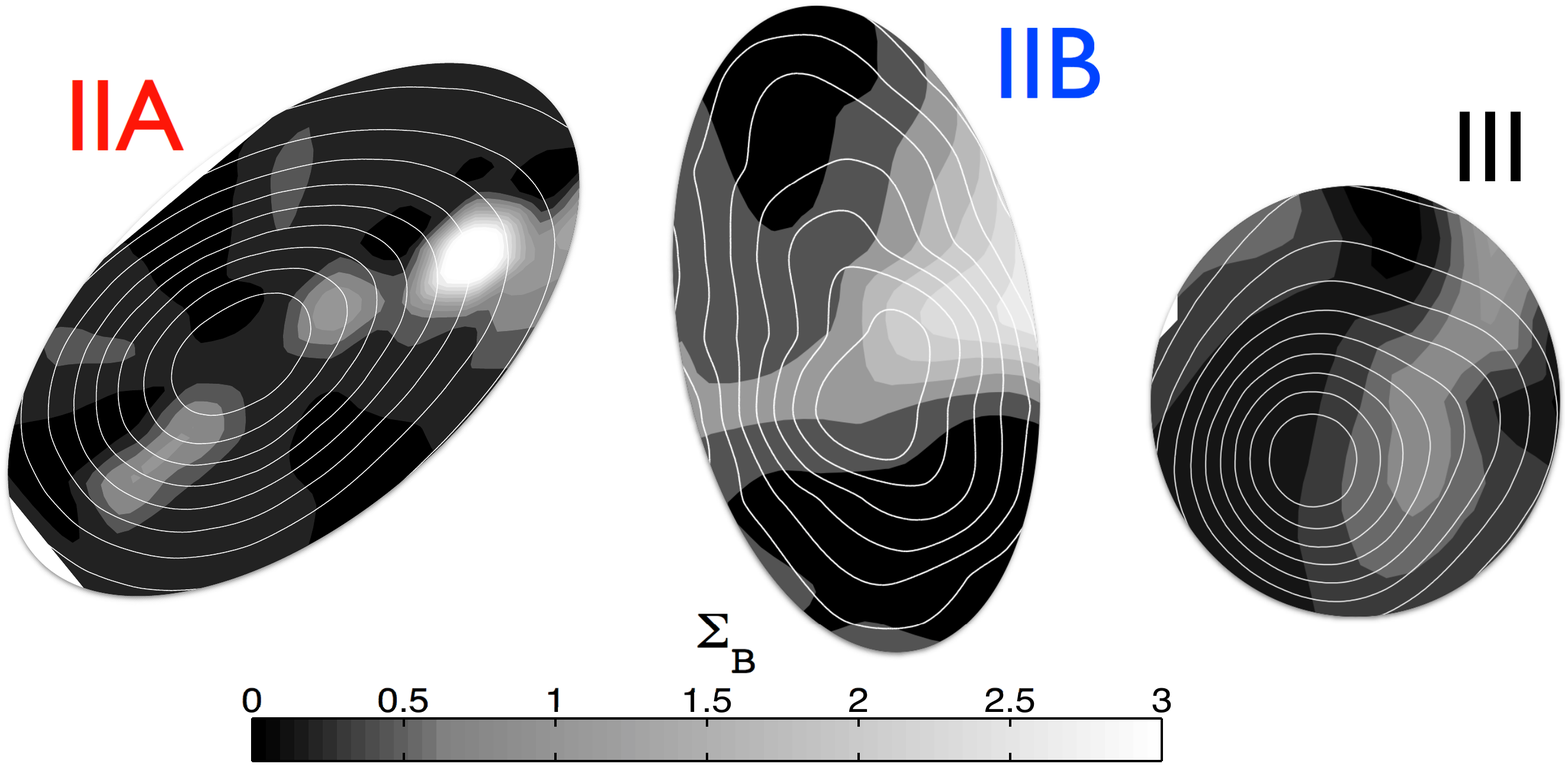}
 \caption{\label{figure_schematic_Sigma_B} 
$\Sigma_B$-maps for the 3 schematic types IIA, IIB and III (Figure \ref{figure_schematic}; 
IIA, IIB: G240 and G35.2 from \citet{zhang14} and \citet{qiu13}; III: W51 e2 from \citet{tang09b}), derived from 
the effectively observed resolutions listed in Table \ref{table_quantities}.
The black-to-white color grading indicates $\Sigma_B$.
Corresponding black and white patterns between $|\delta|$ (Figure \ref{figure_schematic}) 
and $\Sigma_B$ are visible.
}
\end{center}
\end{figure}

It is important to stress that $\langle\Sigma_B\rangle$ describes a source-averaged property. In its
origin, $\Sigma_B$ is, nevertheless, a local measurement. 
$\Sigma_B$-maps (Figure \ref{figure_schematic_Sigma_B} and \citet{koch13})
clearly show distinct zones within the same source where $\Sigma_B$ can be larger or 
smaller than one. Thus, for an average $\langle\Sigma_B\rangle > 1$, local collapse with 
$\Sigma_B < 1$ is still possible. Similarly,  $\langle\Sigma_B\rangle < 1$, does not necessarily
mean that the entire cloud / core is collapsing.

In conclusion, there are clear and systematic differences in $\langle|\delta|\rangle$ across the 
sample of 50 sources. This is entirely based on observed magnetic field and dust morphologies
only. When interpreting $|\delta|$ in the context of the polarization - intensity gradient method, 
a clear correlation is found between $|\delta|$ and the significance of the magnetic field. This 
correlation allows us to discriminate between, on average collapsible and non-collapsible systems.
Despite the fact that the SMA ($\theta \sim 1-3\arcsec$) and the CSO ($\theta \sim 20\arcsec$)
are typically probing different scales and densities, there seems to be no difference between the 
two instruments in Figure \ref{figure_schematic} and \ref{source_Sigma_B_delta}. 
Both the SMA (circles) and the CSO (squares)
equally sample sources along the observed range in $\langle|\delta|\rangle$ and $\langle\Sigma_B\rangle$.
It is only the type-III sources that were exclusively found among the SMA sample (black filled circles).
Further implications of these findings are discussed in Section \ref{discussion}.
Finally, a $3\sigma$-clip is applied in Figure \ref{source_Sigma_B_delta}.
This removes only the largest outliers which are typically found in only 1 or 2 locations in a source.
As a consequence, the spread in 
$\langle\Sigma_B\rangle$ is reduced. A clear trend and distinct regimes are already apparent 
in Figure \ref{figure_schematic} without any clipping.
Table \ref{table_quantities} summarizes these values.

\section{Discussion} \label{discussion}

\subsection{Magnetic Field Significance $\langle\Sigma_B\rangle$ as Measured from $\langle|\delta|\rangle$} 
                                          \label{discussion_fit}

Figure \ref{source_Sigma_B_delta} displays source-averaged force ratios (field tension force
over gravitational force), $\langle\Sigma_B\rangle$, versus source-averaged $\langle|\delta|\rangle$
(field alignment deviation). The clear observed trend leads to a log-linear best-fit with a 
scaling $\langle\Sigma_B\rangle = A\cdot \exp(B\cdot \langle|\delta|\rangle)$ with a mean 
error $d\Sigma_B=\pm 0.20$ and a standard deviation of 0.19 for $A=0.116$ and $B=0.047$.
{\it Can we extract any further meaning from exponent and numerical factor in this scaling?}
The magnetic field significance $\Sigma_B=\sin\psi/\sin\alpha$ is derived in \citet{koch12a}.
$\alpha$ is the complementary angle to the field alignment $|\delta|$, i.e., $\alpha=\pi/2-|\delta|$
(Figure \ref{schematic_simple}).
In the following, we analyze our sample-based scaling for (i) $\langle|\delta|\rangle$ small and 
(ii) $\langle|\delta|\rangle$ large.

For small angles $|\delta|$, we can write $\langle\Sigma_B\rangle=\left\langle\sin\psi/\cos|\delta|\right\rangle
\approx \langle\sin\psi\rangle \approx \langle \psi - \psi^3/3! + \cdots \rangle$. $\psi$ is expressed in 
radians and, if also assumed to be small, it simply becomes $\langle\Sigma_B\rangle \approx \langle \psi \rangle$.
On the other hand, expanding the above scaling, $\langle\Sigma_B\rangle \approx A + A\cdot B \cdot \langle|\delta|\rangle
+ 1/2 \cdot A\cdot B^2 \cdot \langle|\delta|\rangle^2 + \cdots$. Here, $\langle|\delta|\rangle$ is expressed 
in degree, and the zeroth and first term dominate for small $\langle|\delta|\rangle$. Setting equal the 
two equations for $\langle \Sigma_B \rangle$, yields $\langle\psi\rangle \approx A + A\cdot B\cdot \langle|\delta|\rangle$.
The numerical factor $A$, thus, sets the limit for the sample's smallest $\langle \psi \rangle$ if 
$\langle|\delta|\rangle \approx 0$. For the best-fit value $A=0.116$, $\langle \psi \rangle \approx 7^{\circ}$.
The exponent $B$ (together with $A$) defines the slope of how $\langle \psi \rangle$ increases
across the sample with growing $\langle|\delta|\rangle$. As $B$ is very small, $\langle \psi \rangle$ 
increases very slowly. For $\langle|\delta|\rangle = 20^{\circ}$, $\langle \psi \rangle \approx 14^{\circ}$.   
Finding small angles $\langle\psi\rangle$ confirms the initial assumption of $\psi$ being small in the 
above development. 

For large $\langle|\delta|\rangle$ we can write $\langle\Sigma_B\rangle = \left\langle\sin\psi/\cos|\delta|\right\rangle
\approx \sin \psi_s /\langle \cos|\delta|\rangle \approx \sin\psi_s \cdot \langle (1+|\delta|^2/2+ \cdots)\rangle$, 
where we assume $\psi \approx const \approx \psi_s$, but still variable from source to source. Higher-order
terms in $|\delta|$ are negligible as $|\delta|$ is expressed in radians. In the expansion of the 
scaling relation, the linear term is negligible, leading to $\langle\Sigma_B\rangle \approx A + 
1/2\cdot A \cdot B^2 \cdot \langle|\delta|\rangle ^2$. For the numerical values $A$ and $B$, the quadratic 
term further dominates over the zeroth order. Setting equal again the two expressions gives
$1/2\cdot A\cdot B^2 \cdot \langle|\delta|\rangle ^2 \approx \sin \psi_s \cdot \langle (1+|\delta|^2/2)\rangle$. 
For our numerical values and large $|\delta|\approx 60^{\circ}$, this yields $\psi_s\approx 20^{\circ}$.
The angle $\psi_s$ grows, to a first order, with $\langle|\delta|\rangle$ but depends on $B^2=0.05^2$,
which leads again to small values $\langle\psi\rangle$.

\begin{figure}
\begin{center}
\includegraphics[scale=0.7]{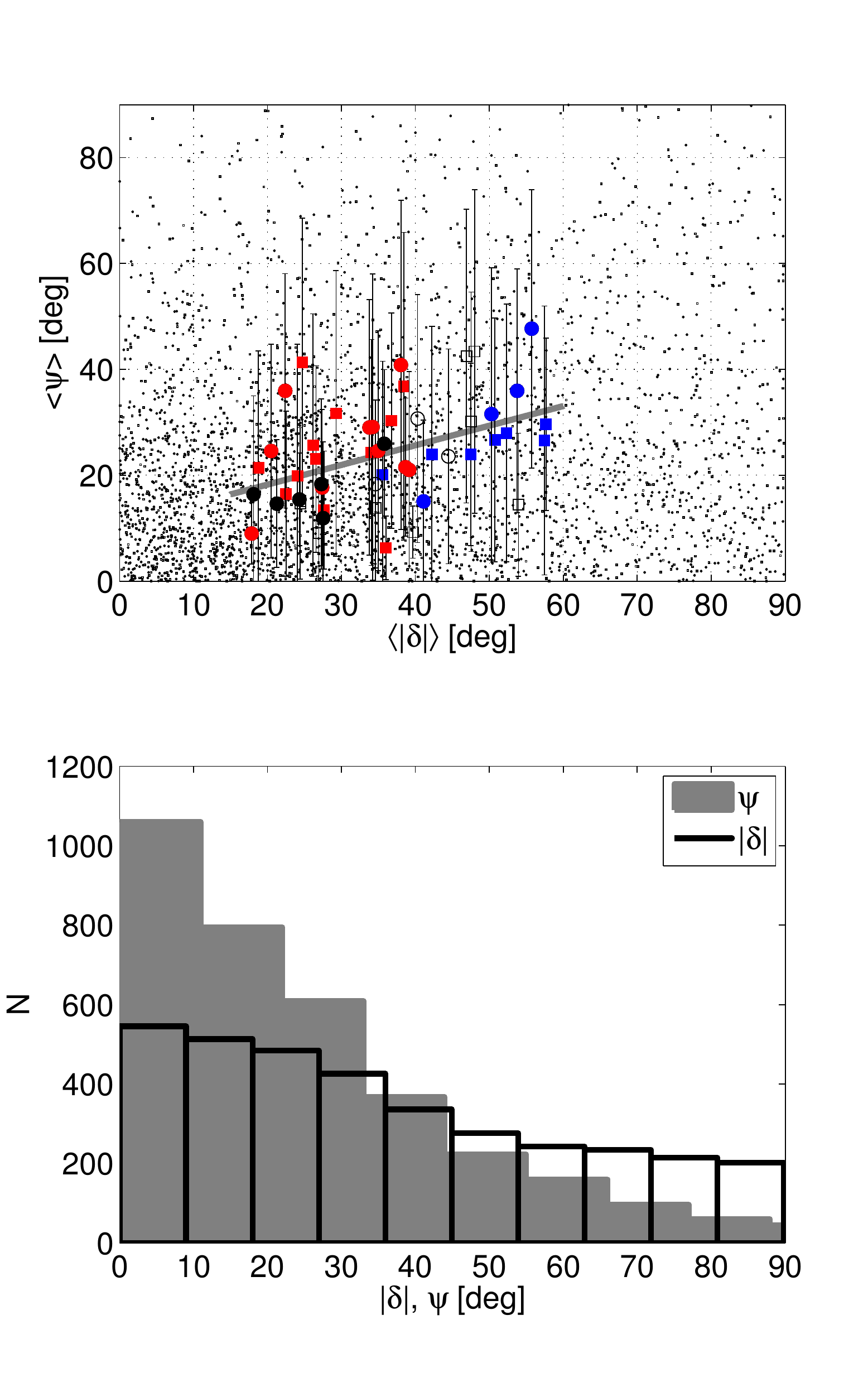}
 \caption{\label{source_psi_delta} \footnotesize
Top panel: Source-averaged $\langle \psi\rangle$-values vs source-averaged 
$\langle |\delta|\rangle$-values for the combined SMA and CSO samples.
Symbols and colors are identical to the ones in Figure \ref{figure_schematic}.
Vertical lines show the dispersion in $\psi$ for each source. The thick gray line
is a linear fit to the averaged values, revealing an only small average increase in 
$\langle\psi\rangle$ with $\langle |\delta|\rangle$ across the entire sample
from $\sim 18^{\circ}$ to $\sim 35^{\circ}$. This is significantly less than 
the change in $\langle |\delta|\rangle$ which makes $\langle |\delta|\rangle$
the main parameter to drive changes in the magnetic field significance $\Sigma_B$.
Small black dots show all independent $(|\delta|,\psi)$ measurement pairs.
Lower panel: Histograms for the SMA-CSO-combined sample for $|\delta|$
and $\psi$ for each measured independent position (not averaged).
While $|\delta|$ shows a flatter distribution, $\psi$ is clearly peaked toward
small angles. This suggests again rather constant, small values for $\psi$ 
with minor changes across the sample, as it is found in the upper panel
for averaged angles.
}
\end{center}
\end{figure}

We briefly turn our attention to the angle $\psi$, the difference between emission gradient and (local)
gravity direction (Figure \ref{schematic_simple}). $\langle\psi\rangle$ is generally found 
to be rather small across the entire 
sample (Figure \ref{source_psi_delta}), slowly growing from an average $18^{\circ}$ to $35^{\circ}$. 
In a perfectly symmetric and circular source, emission gradient and gravity direction 
coincide and $\psi\equiv 0$. More realistically, in close-to-circular, elliptical or still compact
systems, $\psi$ still remains small with little variations across the source (as e.g., in 
W51 e2, IRAS 16293 A, IRAS 16293 B, NGC 1333 IRAS 4A, Orion BN/KL main core, I18360, G34 N, G240...). 
Our analysis of the scaling relation together with Figure \ref{source_psi_delta} possibly indicates a universal
source property with $\langle\psi\rangle$ being generally small. This is likely a general trend 
because we are dealing with confined sources (displayed with closed emission contours) that show 
an accumulation of mass. Overall similar geometries lead to $\psi$ covering similar ranges in each 
source, consequently leading to a similar spread around similar source-averaged $\langle\psi\rangle$
values.

In summary, linking the sample-based scaling of $\langle\Sigma_B\rangle = A\cdot \exp(B\cdot \langle|\delta|\rangle)$
to the original derivation of $\Sigma_B$ shows that $\langle\Sigma_B\rangle$ is largely 
controlled by the source-averaged magnetic field alignment $\langle|\delta|\rangle$. The observed
range for $\langle|\delta|\rangle$ is $\sim 17^{\circ}$ to $60^{\circ}$. For both small and large values
in $\langle|\delta|\rangle$, the average emission gradient to gravity misalignment $\langle\psi\rangle$
is estimated to be similar with $\sim 14^{\circ}$ and $\sim 20^{\circ}$, respectively. 
The really observed range grows from $\sim 18^{\circ}$ to $35^{\circ}$.
Thus, 
$\langle\Sigma_B\rangle$ is indeed influenced to a much smaller extent by $\langle\psi\rangle$.
This further supports the interpretation in \citet{koch13} where, based on the close resemblance
between $|\delta|$- and $\Sigma_B$-maps, $|\delta|$ was proposed to be a tracer for the change in 
the magnetic field significance across an observed source.

\subsection{Observed $|\delta|$ versus Numerical Study}

Morphological imprints in low- to high-magnetization simulated molecular clouds, analyzing relative 
orientations between magnetic field and column density gradients, were studied in \citet{soler13}.
Their 2D-projected analysis reveals differences in distributions of their angle $\phi$ (between 
magnetic field and column density gradients) as a function of magnetization, column density and 
evolutionary stage. While histograms for low-magnetization cases (weak field strength) peak for all 
densities and time steps around $\phi\sim 90^{\circ}$, indicating magnetic fields that are 
tracing iso-contours in column densities, high-magnetization cases (large field strength) display 
a shift in distribution peaks as column densities increase from magnetic fields tracing iso-contours
to magnetic fields perpendicular to contours (Figure 9 in \citet{soler13}). Such opposite trends are 
precisely seen in our $\delta$-histograms for IIA- vs IIB-type sources (Figure \ref{hist_cat}). We note 
that the angle $\phi$ in \citet{soler13} is defined from 0 to 180$^{\circ}$ where $\phi=90^{\circ}$
indicates a magnetic field orientation perpendicular to a column density gradient. We have defined 
$\delta \in [-90, 90]$ where $\delta = 0^{\circ}$ means the field is parallel to an emission gradient
and $\pm 90$ indicates counter-clock or clockwise rotation. $\phi$ peaking around 90$^{\circ}$ is, thus, 
equivalent to $\delta$ peaking around $\pm 90^{\circ}$ (IIB), and $\phi$ peaking around 0 and 180$^{\circ}$
corresponds to a peak in our $\delta$-histograms around 0 (IIA). In the framework of the simulations 
by \citet{soler13}, this indicates that IIA sources in both the CSO and the SMA samples are 
expected to have strongest magnetic fields and largest column densities. This will even more 
apply to the SMA type-III sources because they show an even more pronounced peak around $\delta\sim 0$.
Indeed, large field strength values from a few mG to $\sim 10$~mG are measured in more evolved, 
collapsing type-III sources for W51 e2 \citep{tang09b, koch12a}, G31.41+0.31 \citep{girart09} and 
NGC 1333 IRAS 4A \citep{girart06}. It needs to be mentioned that IIA/III-type sources having 
the largest field strengths does not contradict with our finding in Figure \ref{source_Sigma_B_delta}
where IIA/III-type sources show $\langle \Sigma_B \rangle \simless 1$. $\Sigma_B$ being a ratio, 
a field strength can significantly increase as material is compressed while, at the same time, 
a growing gravitational pull from a denser core can still lead to $\Sigma_B \simless 1$. The 
detailed analysis in W51 e2 \citep{koch12b} is supporting this picture, revealing growing 
field strength values from a few mG to about 17~mG in the inner core while $\Sigma_B$ is
indeed decreasing toward the center.

A source-by-source comparison in our sample with the results in \citet{soler13} is non-trivial
because individual sources often do not have enough independent detected polarization segments
in order to unambiguously categorize a $\delta$-histogram according to their classification scheme.
We have, therefore, focused on the clearly identified groups IIA, IIB and III in our data sets. 
By accumulating distributions in this way, we are able to identify differences in $\delta$-distributions
that can be matched with numerical results.

\subsection{Magnetic Field and Source Orientation}

\subsubsection{Previous Observational Studies}

\citet{tassis09} analyzed a sample of 24 molecular clouds, observed in submillimeter dust emission 
at 350 $\mu$m with the CSO/Hertz with a resolution of about 20$\arcsec$ \citep{dotson10}. Their 
study is motivated by the goal to determine the mean orientation of the ordered magnetic field with 
respect to the cloud principal axes. In their approach they compare the mean dust polarization 
orientation (derived from all the measured polarization position angles in a cloud) with the cloud
aspect ratio (derived from the first and second moments of the dust emission). A statistical 
treatment, assuming the orientations of the molecular clouds to be random with respect to the 
line of sight and equally probable, is adopted to overcome projection effects. \citet{tassis09}
find that oblate cloud shapes are the most likely geometry for their CSO sample while the magnetic
field shows a preference to be aligned with the shortest cloud ellipsoidal axis with a deviation
of $\sim 24^{\circ}$ towards the middle and long cloud axes. While their best-fitting distribution 
is peaking around this value, they note that long tail features are present, and the relatively 
small number statistics still sets limits to their conclusion. They rule out a scenario where the 
magnetic field orientations are drawn randomly from a uniform distribution at the percent level.

In a recent study, \citet{li13} investigated the orientations of 13 filamentary molecular clouds
in the Gould Belt with respect to the intercloud medium (ICM) using near-infrared dust extinction
maps and optical stellar polarimetry data. These filaments have low column densities 
($\langle A_v \rangle < 2$~mag) with length scales of $\sim 1-10$~pc. \citet{li13} define the 
orientation of a cloud by the long-axis orientation of the auto-correlation function of the column
density map. In order to derive the mean magnetic field orientation of the surrounding ICM (outside
of the cloud), an equal-weight Stokes mean of all the detected polarization is adopted. They find 
that all mean field and cloud orientations fall within $30^{\circ}$ from being either perpendicular
(7 clouds) or parallel (6 clouds) to each other. They exclude the possibility that such a sample
can be drawn from random orientations with a probability of less than 1\%. \citet{li13} conclude
that their finding of a bimodal distribution supports the scenario of a dynamically important
magnetic field, disfavoring super-Alfv\'enic turbulence that would lead to random magnetic field
orientations with clouds \citep[e.g.,][]{padoan01}. 

A sample of 14 high-mass star-forming clouds, observed with the SMA in dust polarization at 
870$\mu$m with a resolution of a few arcseconds, serves as a basis of a statistical analysis
in \citet{zhang14}. The SMA sample typically probes scales around 0.1 to 0.01~pc and densities
larger than $10^5$cm$^{-3}$. \citet{zhang14} determine the position angle of the clump major
axis by fitting a 2-dimensional elliptical Gaussian to the dust continuum emission. The magnetic 
field orientation over the clump is not averaged but rather for every independent polarization 
measurement, the absolute difference in angle between the clump major axis and every measured
polarization segment is computed. The histogram of these angles reveals a bimodal distribution
with two peaks: a first population of magnetic field orientations that is aligned with the clump 
major axis within about 10$^{\circ}$, and a second population associated with magnetic field 
segments that are perpendicular to the clump major axis within about $40^{\circ}$.

Besides the above summarized studies where polarization is detected across or outside of each cloud or clump
with at least several or many independent measurements, earlier studies already probed magnetic
field orientations but were limited to a single detection of polarized orientation at the source
intensity peak. These studies are summarized in \citet{tassis09}. The results vary from apparent
random orientations with respect to cloud axes (\citet{glenn99}, 7 elongated cores) to a strong
preference of polarized emission aligned with structures (\citet{kane93}, 6 sources), and magnetic field
orientations in cloud cores to be parallel in 3 out of 10 cases \citep{vallee99}.

\subsubsection{$|\delta|$ and Bimodality in Magnetic Field versus Source Orientation?}

{\it How do our results for the angle $|\delta|$ (Figures \ref{hist_cat} and \ref{source_Sigma_B_delta}) 
compare to all the previous studies?} The first point to note is that all the earlier work is 
dealing with samples of only a few up to about 10 sources, with the largest sample being 24 sources 
in \citet{tassis09}. Our analysis is based on 50 sources. A second major difference is
the analyses and steps taken in order to arrive at the conclusion of aligned or perpendicular 
fields with respect to a cloud or a core. Essentially every cited work is taking a different 
approach. This also includes our work here. Further relevant to these different approaches is the 
concept of {\it average} (one single global) quantity versus {\it local} quantities. \citet{tassis09}
and \citet{li13} compare a single average magnetic field orientation (based on all measured 
polarization P.A.s and equal-weight Stokes mean of all detected polarization, respectively) with a 
single average structure orientation (derived from moments aspect ratios and column density
auto-correlations, respectively). \citet{zhang14} compare a single average source orientation 
(based on 2-dimensional ellipsoidal fitting) with every independent local magnetic field orientation.
Our work here compares every independent local field with every independent local emission gradient
orientation. Not only differ all these studies in what they are comparing (average or local quantities),
but they also define {\it average} in different ways. While different data sets can be suggestive
for different approaches and different authors can have different preferences, this obviously can
introduce biases leading to a non-trivial comparison. 

The starting motivations in \citet{koch12a} to work with {\it local} quantities was that (1) the 
role (and strength) of the magnetic field will generally vary with position in a source; and (2)
averaging might not always be an adequate operation to characterize certain sources. To be more 
explicit, for close-to-point symmetrical or close-to-radial field morphologies, as e.g., in 
W51 e2 \citep{tang09b}, defining an average field orientation is not unambiguous. Such systems
would typically allow for two possible average orientations with a 90$^{\circ}$ difference between
them. In practise, sources will not be perfectly symmetric. Non-symmetric features, especially if 
further weighted by, e.g., Stokes $I$, can then tip the balance in favor of one or the other 
possible average orientation. It is, however, questionable, whether such a subtlety is physically 
motivated. It is unlikely to provide a robust criterion for such sources. Mathematically, the
reason for this ambiguity and complication is the large dispersion between field P.A.s. This is 
unlike sources that show close-to-uniform field orientations that lead to a small dispersion.
A second group of sources where averaging is not well motivated (although it can be computed), 
are complicated systems, such as, e.g., DR 21(OH) \citep{girart13} where fragmentation
and multiple outflows are leading to complex (but still organized) field morphologies. 
Obviously, one could remedy this situation by simply excluding any questionable sources from 
any statistical survey studies. We are refraining from that because our goal is to provide
a framework to assess the role of the magnetic field in any source. Based on this reasoning, 
our analysis here is adopting the local quantity $|\delta|$\footnote{
An additional property of $\delta$ is that it is unambiguously defined in the range -90
to $+90^{\circ}$ (if giving it a sense of orientation) or 0 to $+90^{\circ}$ (if working with 
absolute values). This is because $\delta$ and $|\delta|$ are {\it differences} between 2
P.A.s which makes them rotationally invariant quantities, i.e., they are independent of any 
reference coordinate. This is unlike P.A.s that are measured with respect to some orientation 
and reference frame. Additionally, P.A.s suffer from a 180$^{\circ}$-ambiguity resulting from the
fact that we are measuring orientations only and not directions with vector heads. 
}.
Despite the differences and complications in comparing the various studies, we are now turning
our attention to the question whether there is bimodality in $|\delta|$ or $\langle|\delta|\rangle$
(the source-averaged quantity displayed in Figure \ref{source_Sigma_B_delta}).

\begin{figure}
\begin{center}
\includegraphics[scale=0.7]{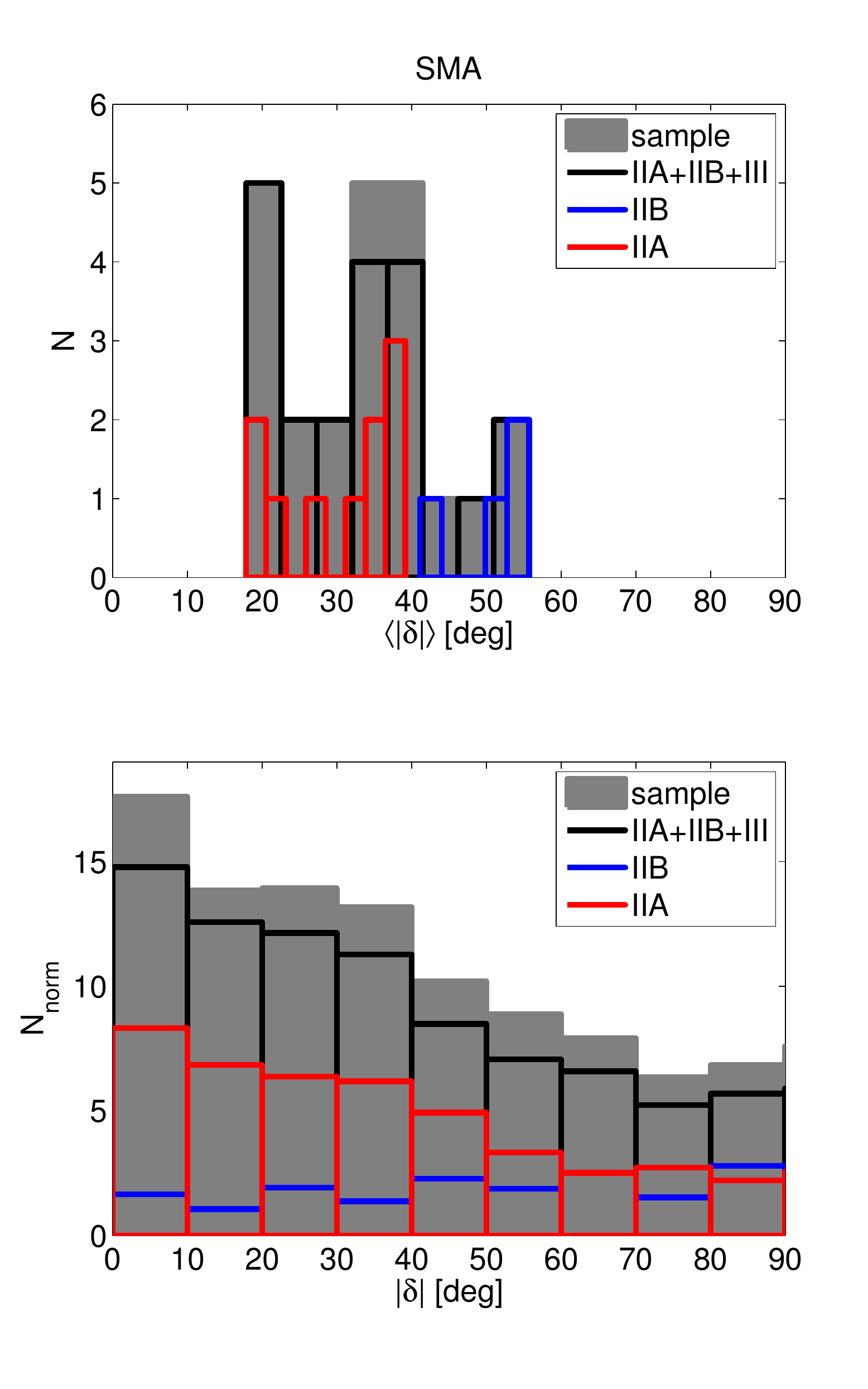}
 \caption{\label{bimodality_sma} \footnotesize
Histograms for the SMA sample: source-averaged $\langle|\delta|\rangle$ (top panel)
and $|\delta|$ (bottom panel). Bins are slightly shifted in the upper panel for a better
display of the various groups. {\it 'Sample'} refers to the entire sample (23 sources) 
including sources that are not categorized. In the bottom panel, $|\delta|$-histograms
for each source are normalized to 1, and then summed up in each bin.
}
\end{center}
\end{figure}

\begin{figure}
\begin{center}
\includegraphics[scale=0.7]{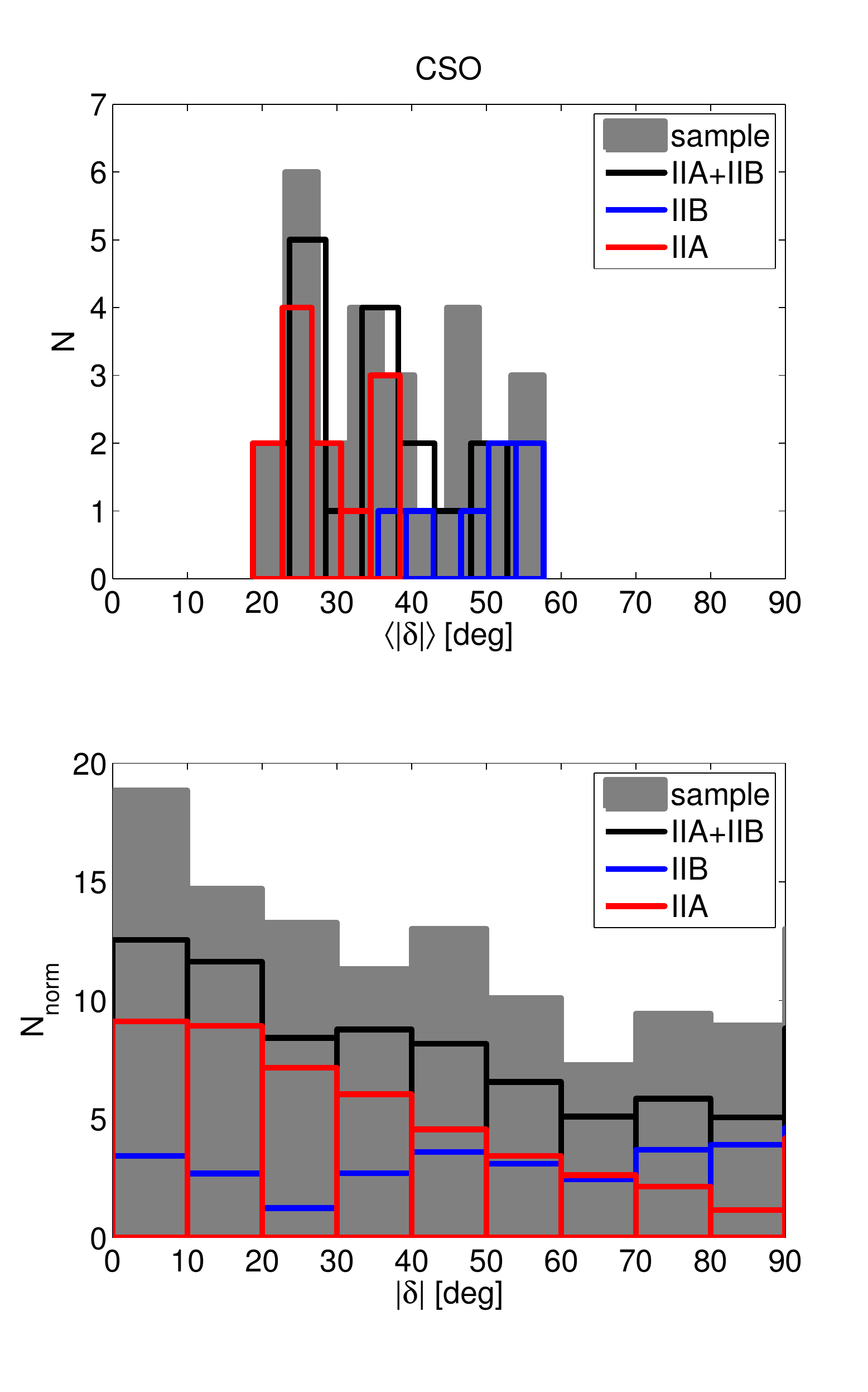}
 \caption{\label{bimodality_cso} \footnotesize
Identical to Figure \ref{bimodality_sma} but for the CSO sample. 
No type-III sources are identified in this sample. 
}
\end{center}
\end{figure}

Histograms for $\langle|\delta|\rangle$ and $|\delta|$ are shown in Figure \ref{bimodality_sma}
for the entire SMA sample and its different categories IIA, IIB and III. As expected, IIA and 
IIB populate the lower and upper end in the $\langle|\delta|\rangle$-histogram. Type-III sources
mostly add to the lower end. Interestingly, some IIA and III sources also fall around 
$\langle|\delta|\rangle\approx 40^{\circ}$. The remaining sources -- neither classified as
IIA, IIB nor III -- exclusively fall around mid-values $\sim 40^{\circ}$. We remark again that the 
categorization of IIA, IIB and III is simply motivated by visual inspection and does in no way
affect the derivation of $\langle|\delta|\rangle$. While there is an indication of a possible
bimodality with peaks around $\sim 20^{\circ}$ and  $\sim 55^{\circ}$ -- driven mostly by IIA
and IIB sources -- the histogram reveals a third peak around $\sim 40^{\circ}$.
{\it What causes this central peak?} 
$|\delta|$ very clearly and effectively differentiates between magnetic fields locally 
aligned or perpendicular to a source structure (emission gradient). Consequently, when 
calculating an average $\langle|\delta|\rangle$, larger or smaller source major-to-minor axes ratios can 
lead to a spread around typical IIA and IIB $\langle|\delta|\rangle$-values; e.g., for a source 
with an axes ratio 3:1 with a magnetic field parallel to its minor axis (type IIA), 
$\langle|\delta|\rangle$ will be small $\sim 20^{\circ}$. If the source is less filamentary, 
becoming more compact and shorter (but still with the field parallel to its minor axis), the 
axes ratio decreases and $\langle|\delta|\rangle$  will grow to $\sim 30-40^{\circ}$.
Similarly, for IIB-sources where the field is aligned with the source major axis, 
$\langle|\delta|\rangle$ will decrease as the source becomes less filamentary. All this is 
simply a result of averaging over more or fewer small (IIA) or large (IIB) $|\delta|$-values
because of more or less elongated sources. This then explains a spread in $\langle|\delta|\rangle$, 
that can populate mid-values. Studies that are comparing mean source orientations versus mean or 
individual magnetic field orientations \citep{tassis09,li13,zhang14} are less subject to such a spread.
The second group falling onto mid-$\langle|\delta|\rangle$ values are sources that are neither
categorized as IIA, IIB nor III. Although there are few of these sources, they do exist.
In summary, lower- and upper-end peaks in the top panel in Figure \ref{bimodality_sma} can become more 
pronounced by taking into account spreading in $\langle|\delta|\rangle$ for filamentary sources.
This would point toward bimodality. At the same time, nevertheless, there are clearly sources
that fall in between. The histogram for the CSO sample reveals a very similar result
(Figure \ref{bimodality_cso}, top panel), with an upper-end peak that appears more present. 

\begin{figure}
\begin{center}
\includegraphics[scale=0.7]{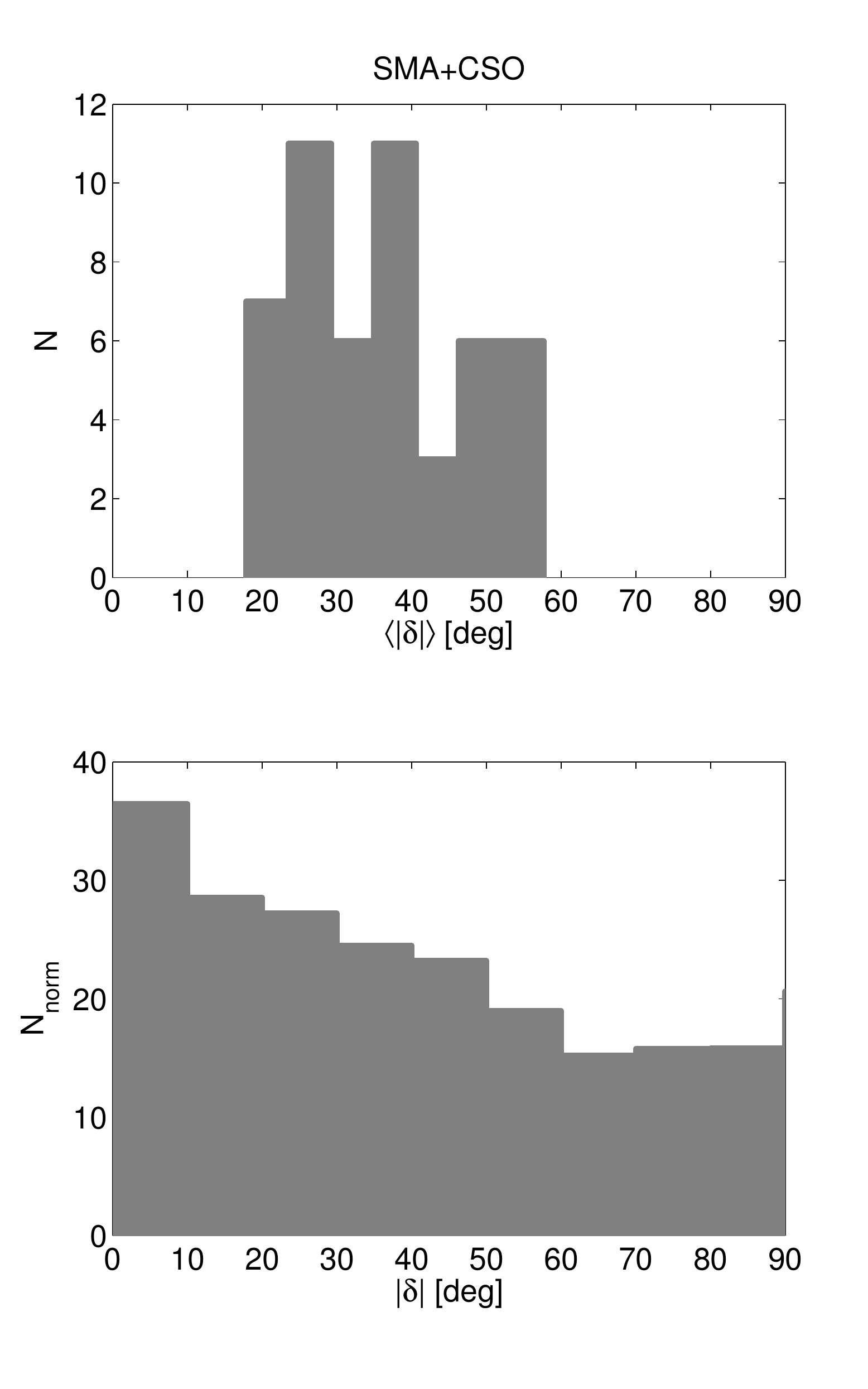}
 \caption{\label{bimodality_sma_cso_combined} \footnotesize
Identical to Figure \ref{bimodality_sma} but for the SMA-CSO-combined sample. 
}
\end{center}
\end{figure}

{\it Is there bimodality in $|\delta|$ before averaging?} The bottom panels in the Figures
\ref{bimodality_sma} and \ref{bimodality_cso} show normalized histograms where $|\delta|$-histograms
for each source are normalized to one and then summed up. This normalization removes
a possible bias where (nearby) sources with many more independent measurements could dominate
the statistics. IIA- and IIB-categories show the expected trends with falling and rising
distributions with larger $|\delta|$. This is largely consistent with the histograms in 
Figure \ref{hist_cat} that are counting every single independent measurement. Combining the 
two categories results in a distribution falling with $|\delta|$. This is further amplified 
when adding type-III sources for the SMA and non-categorized sources. 
Assuming that the SMA and CSO samples represent a fair cross-section of star-forming
regions, {\it this result shows that the prevailing source-magnetic field configuration
in both the SMA and CSO sample is one that statistically prefers small magnetic field misalignments $|\delta|$.}
With this additional information we might re-interpret the upper panels in Figure 
\ref{bimodality_sma} and \ref{bimodality_cso}: the $\langle|\delta|\rangle$-histograms
possibly hint a maximum peak around smallest $\langle|\delta|\rangle$-values with a tail toward
larger $\langle|\delta|\rangle$-values. This interpretation is gaining additional support
when combining the SMA and CSO samples into one $\langle|\delta|\rangle$-histogram 
(Figure \ref{bimodality_sma_cso_combined}), which shows a broader distribution. 
$\langle|\delta|\rangle$ is less than $45^{\circ}$ in 75\% of all the sources. One third of the 
sources shows $\langle|\delta|\rangle < 30^{\circ}$.
A uniform distribution seems to be ruled out.

We finally note that earlier studies might already have seen hints of such distributions as well.
\citet{zhang14} suggest a combination of two distributions (0$^{\circ}$ to 40$^{\circ}$ and 
80$^{\circ}$ to 90$^{\circ}$) with a weight 5:3 to explain their histogram measuring angles
between source major axis and polarization orientations. Both \citet{zhang14} and \citet{li13}
state bimodality based on 'fields aligned within 40$^{\circ}$' and 'orientations being 
perpendicular within 30$^{\circ}$'. These are wide ranges, given only a 0$^{\circ}$ to 90$^{\circ}$
range to fully account for between perfectly aligned and perfectly perpendicular. Bimodality, if at all,  
is, thus, very unlikely to occur with sharp and narrow peaks. With the to date largest 
sample of 50 sources it seems to become more plausible that we are dealing with broad
distributions peaking at small misalignment values with a tail and a possibly secondary
peak or bump toward larger misalignments. With a sample of 24 sources, \citet{tassis09} 
also noticed the presence of long-tail 
features in their analysis. If further confirmed with even larger samples, this finding would
then favor a scenario where the dominating configuration in sources from $\sim$pc to $\sim$ mpc
scale has the magnetic field roughly aligned with the source minor axis.

On much larger scales ($\sim 10\arcmin$), tracing the interstellar medium (ISM)
         with polarized dust emission at 353~ GHz, recent {\it Planck} all-sky observations
          \citep{planckxxxii},
         investigating the relative orientations between the magnetic field and dust
         structures, show that in the diffuse ISM structures are preferentially aligned with the 
         magnetic field. A glimpse into denser molecular cloud complexes further reveals that
         this degree of alignment decreases with increasing column density. The magnetic 
         field tends to become more perpendicular to the interstellar dust ridges, i.e., aligned
         with their shorter axes. Our results for $|\delta|$ on smaller scales and higher densities
         agree with this trend.

\subsection{Reduced Star-Formation Efficiency and Mass-to-Flux Ratio}

\begin{figure}
\begin{center}
\includegraphics[scale=0.7]{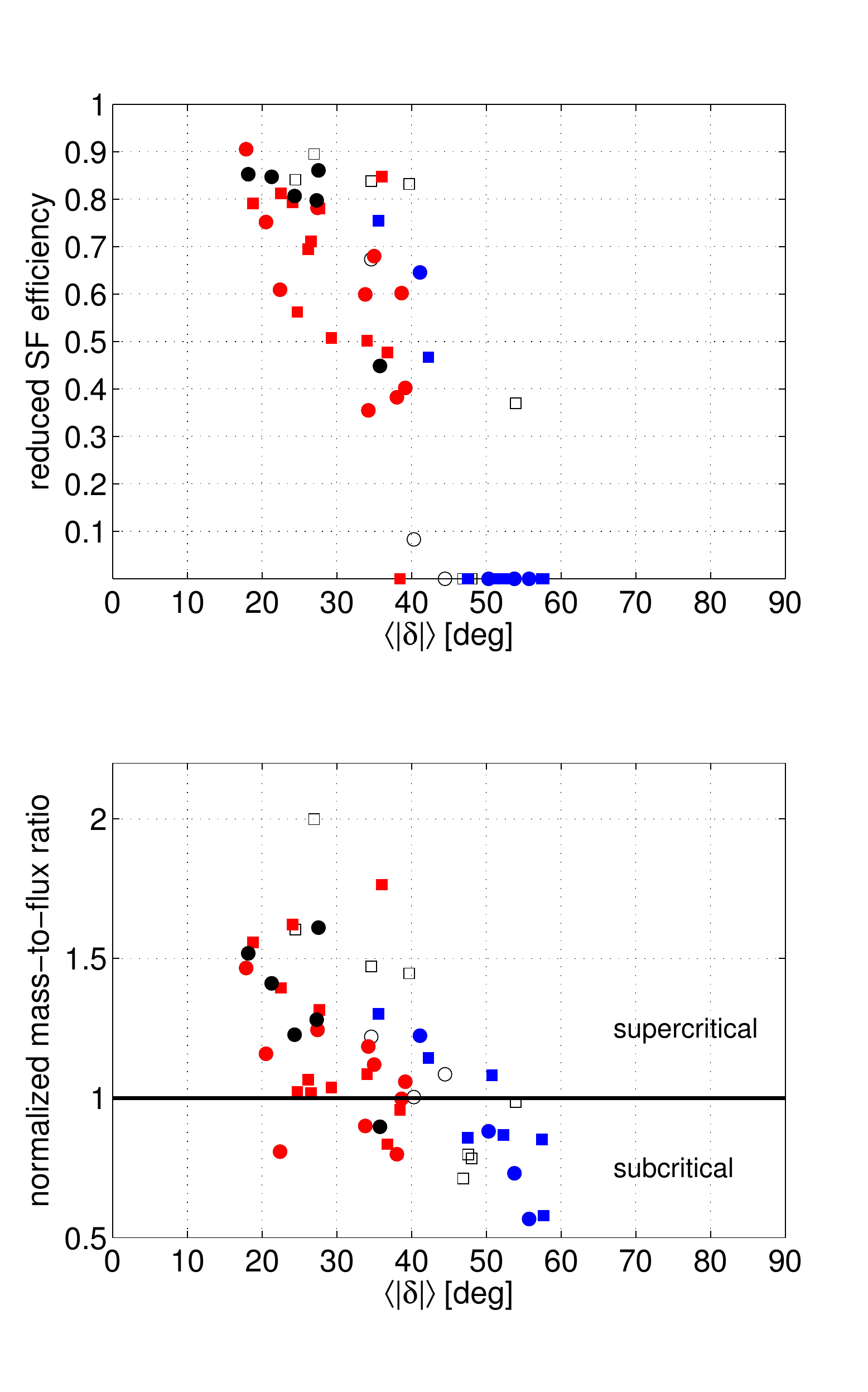}
 \caption{\label{efficiency_mass_flux} \footnotesize
Magnetic-field-reduced star formation efficiency (top panel) and 
normalized mass-to-flux ratio (bottom panel) as a function of 
source-averaged $\langle|\delta|\rangle$. The solid black line
marks the critical mass-to-flux ratio, separating 
systems that are, {\it on average}, supercritical or subcritical.
In either case, {\it local collapse} can be hindered or possible according
to the schematic $|\delta|$- and $\Sigma_B-$patterns in Figure \ref{figure_schematic}
and \ref{figure_schematic_Sigma_B}.
Symbols and colors are identical to the ones in Figure \ref{figure_schematic}.
}
\end{center}
\end{figure}

Having derived a field significance $\langle\Sigma_B\rangle$ for each source in 
Figure \ref{figure_schematic} and \ref{source_Sigma_B_delta}, 
we can now proceed to estimate further source properties: 
star-formation efficiency and mass-to-flux ratio. Since the magnetic field is providing
resistance against gravity -- quantified with the magnetic field tension-to-gravity
force ratio $\Sigma_B$ -- gravity is not everywhere equally effective to initiate and drive
a collapse. A concept of magnetic-field-diluted gravity was explored in \citet{shu97}.
With $\Sigma_B$ we can now define a {\it gravity efficiency} $\epsilon_G\equiv 1- \Sigma_B$;
i.e., if $\Sigma_B=0$ or very small, the entire non-diluted gravitational force from an 
enclosed mass distribution is driving a collapse and the gravity efficiency $\epsilon_G \approx 1$;
if $\Sigma_B$ is large, the magnetic field is holding against gravity and the effectively 
acting gravity appears reduced with $\epsilon_G < 1$ \citep{koch12b}. The gravity efficiency
is limited to the range between 0 and 1. We set $\epsilon \equiv 0$ if $\Sigma_B >1$
where collapse is inhibited. The free-fall star-formation efficiency is defined
from the free-fall collapse time and mass accretion solving a (pressureless) momentum
equation with only a gravitational force term. Absorbing the gravity efficiency factor 
$\epsilon_G$ into an effective gravitational force leads to a magnetic-field-reduced
star-formation efficiency (as compared to a free-fall efficiency) $\propto \epsilon_G^{1/2}$
where the square root appears because collapse time is inverse proportional to the 
square root of the enclosed mass \citep{koch12b}. Figure \ref{efficiency_mass_flux}
displays magnetic-field-reduced star-formation efficiencies,
$\langle\epsilon_G^{1/2}\rangle=\sqrt{1-\langle\Sigma_B\rangle}$, for the entire 
SMA and CSO samples\footnote{
Here, we are looking for sample-wide trends and, therefore, adopt a source-averaged
gravity efficiency $\langle\epsilon_G\rangle$ based on $\langle\Sigma_B\rangle$.
It is, nevertheless, worth noting that $\epsilon_G$ like the prime observable $\delta$
can be locally measured. In \citet{koch12b}, azimuthally averaged gravitational efficiencies
$\epsilon_G(r)$ showed clear radial profiles for W51 e2/e8 from several observations, being
close to 1 in the core and dropping to $\sim 20$\% or less in the outer regions.
} as a function of $\langle|\delta|\rangle$.
The sample-wide trend shows that the magnetic field is able to reduce the 
star-formation efficiency. Moreover, the efficiency tends to drop with larger $\langle|\delta|\rangle$.
The SMA type-III sources show a close-to-maximum efficiency, largely unaffected
by the magnetic field. Generally, type-IIB sources reveal a more significantly reduced average
efficiency than IIA sources, dropping to 0 ($\langle\Sigma_B\rangle >1$) and possibly preventing
(further) collapse for $\langle|\delta|\rangle \simgt 45^{\circ}$. 
It is important to note that {\it local} collapse and {\it local} star formation 
are still possible or can be hindered
according to the $\Sigma_B$-patterns in Figure 6.
Sample-averaged efficiencies
are around 50\%. Finally, it needs to be noted that these reduced efficiencies
are upper limits because not all of a cloud's / core's mass might take part in a collapse.
This leads to an additional volume efficiency factor, $\epsilon_V$, which can further 
reduce the total efficiency as $\sim \epsilon_V \cdot \epsilon_G^{1/2}$ \citep{koch12b}.
The conservative estimate here is using $\epsilon_V\equiv 1$.

The second source property we are investigating is the mass-to-flux ratio. Unlike earlier
studies (e.g., \citet{crutcher12} and references therein) that are relying on a combination of 
magnetic field strength measurements (like Zeeman line splitting) and a mass proxy
to estimate this quantity, we are again making use of the field tension-to-gravity force
ratio $\Sigma_B$. Explicitly writing out the forces and expressing them in terms of
magnetic flux and mass enables us to link $\Sigma_B$ to the mass-to-flux ratio.
The source-averaged mass-to-flux ratio, normalized to the critical mass-to-flux 
ratio, can be written as $\langle M/\Phi\rangle_{norm} = \langle\Sigma_B^{-1/2}\rangle
\cdot \pi^{-1/2}$ \citep{koch12b}. The inverse square root results from the magnetic 
flux $\Phi$ being proportional to the field strength $B$ whereas the tension force
is proportional to $1/B^2$. The numerical factor $\pi^{-1/2}$ appears because magnetic
flux is defined across a cylindrical cross-section $\pi R^2$. Normalized mass-to-flux 
ratios\footnote{Analogously to $\delta$ and the gravity efficiency $\epsilon_G$, the 
mass-to-flux ratio based on $\Sigma_B$ can also be locally computed. A clear transition
from magnetically subcritical at larger radii to supercritical towards the core center
was found for W51 e2 in \citet{koch12b}. Here, we are again looking for sample-wide
trends adopting source-averaged mass-to-flux ratios.
} are depicted in the bottom panel in Figure \ref{efficiency_mass_flux} versus 
$\langle|\delta|\rangle$. The overall result is a mass-to-flux ratio decreasing with 
larger misalignment angles $\langle|\delta|\rangle$, leaving most of the type-IIA and 
-III sources in a magnetically supercritical state while most of the type-IIB sources 
fall below the critical mass-to-flux ratio into a subcritical state. This trend and 
transition are largely consistent with the scaling of the field significance $\Sigma_B$
in Figure \ref{source_Sigma_B_delta}.

\section{Summary and Conclusion}  \label{section_conclusion}

A sample of 50 star-forming regions observed in submillimeter dust polarization 
with the SMA and the CSO is studied with the goal of assessing the significance of the 
magnetic field. Our analysis is making use of the fundamental observable $\delta$, 
which is the angle between magnetic field orientation and dust emission gradient. 
The key results are summarized in the following.

\begin{enumerate}

\item 
Across the entire SMA-CSO-combined sample, $|\delta|$ (from about $4,000$ independent measurements)
is populating the entire range between 0 to $90^{\circ}$ with a broad peak 
around smaller values;
i.e., the magnetic field 
is found to range from locally aligned ($0^{\circ}$) to locally maximally misaligned ($90^{\circ}$)
with respect to the dust emission gradient. 
For this large sample, projection effects likely average out. 
When interpreting $|\delta|$ with the polarization--
intensity gradient method, growing $|\delta|$ generally indicates growing magnetic field 
significance $\Sigma_B$ (the ratio of magnetic field tension force to gravity and/or pressure
gradient). For local independent measurements, $\Sigma_B$ can be estimated on average with
$\Sigma_B\approx 1/2\cdot 1/\sin(\pi/2 -|\delta|)$.

\item {\it Categorization:}
Various groups (IIA: tendency of magnetic field to be parallel to source minor axis; IIB: 
tendency of magnetic field to be parallel to source major axis; III: symmetrized source)
show striking differences in group-based $|\delta|$-histograms. Type-IIA and IIB samples
depict opposite trends, peaking at small and large $|\delta|$-values, respectively. Type-III
sources (only found in SMA observations) peak at even smaller $|\delta|$-values.
Sources that can not be categorized, can fall into or in between any of these groups.
Group-based differences in $|\delta|$ manifest themselves with clear differences in the 
group-based field significance: $\langle\Sigma_B\rangle_{IIA},\langle\Sigma_B\rangle_{III}< 1$ 
and $\langle\Sigma_B\rangle_{IIB}>1$.

\item{\it Source-averaged scaling across entire sample:}
A clear correlation is found between the source-averaged field misalignment $\langle|\delta|\rangle$
and the source-averaged field significance $\langle\Sigma_B\rangle$. The combined SMA-CSO
sample yields a best-fit log-linear scaling $\langle\Sigma_B\rangle = 0.116 \cdot \exp(0.047\cdot \langle|\delta|\rangle)$
with a mean error $d\Sigma_B=\pm0.20$. The angle $\psi$ between dust emission gradient and 
local gravity and/or pressure gradient, which is used in the exact derivation of 
$\Sigma_B=\sin\psi/\sin\alpha$, shows only small variations across the sample. This 
explains the tight correlation between $\langle|\delta|\rangle$ and $\langle\Sigma_B\rangle$
and makes $\langle|\delta|\rangle$  a prime observable to trace and quantify the role of the 
magnetic field. These findings also hold separately for both the SMA and the CSO sources, 
possibly hinting a scale-free self-similar magnetic field influence. 
Individual sources that are clearly categorized as IIA (and III) and IIB almost exclusively
show average $\langle\Sigma_B\rangle< 1$ (collapse possible, on average) and 
$\langle\Sigma_B\rangle>1$ (no collapse, on average).
In either case, local collapse can be possible or hindered as $\Sigma_B$-maps show
distinct zones with $\Sigma_B$ larger or smaller than one.

\item{\it Comparison with numerical study:}
Numerical simulations of low- to high-magnetization molecular clouds find distributions in 
$\delta$ that can be matched with our observed IIA- and IIB-type segregation. In this 
framework, IIA- (and III-)type sources represent higher magnetization scenarios with stronger
magnetic fields. Large magnetic field strengths are, indeed, measured for, e.g., W51 e2, 
G31.41+0.31 and NGC 1333 IRAS 4A which are categorized as IIA and III. 

\item{\it Bimodality:}
Distributions of the average magnetic field misalignment $\langle|\delta|\rangle$ for the SMA
and CSO samples possibly are bimodal. However, combining them together broadens the distributions
with a broad maximum peak around small $\langle|\delta|\rangle$-values, a populated mid-range and 
a secondary smaller peak at larger $\langle|\delta|\rangle$-values. Assuming the SMA and CSO samples to be
representative, the most abundant magnetic field-source configuration has, on average, the magnetic field 
roughly aligned with the emission gradient. This is favoring magnetic fields that are roughly aligned 
with the source minor axis. 75\% of the sources show an average misalignment $\langle|\delta|\rangle$
of less than $45^{\circ}$.

\item{\it Star-formation efficiency and mass-to-flux ratio:} 
Based on the field tension-to-gravity force ratio $\Sigma_B$, an effectively acting 
gravitational force can be defined with a gravity efficiency factor $\epsilon_G \equiv 1-\Sigma_B\le 1$.
Consequently, in the presence of a magnetic field, the standard free-fall collapse time grows
and the star-formation efficiency is reduced. The sample-wide trend shows a drop in star-formation 
efficiency with larger $\langle|\delta|\rangle$ with the average star-formation efficiency being significantly reduced or even 
suppressed for some sources with $\langle|\delta|\rangle \simgt 45^{\circ}$.
$\Sigma_B$ can also be linked 
to the normalized mass-to-flux ratio. A transition from magnetically supercritical (mostly for 
type-IIA and -III sources with small $\langle|\delta|\rangle$) to subcritical (for sources with 
larger $\langle|\delta|\rangle$) is found across the sample.

\end{enumerate}


The authors thank the referee for valuable comments and suggestions that further improved 
this manuscript. 
SPL and TCC are supported by the Ministry of Science and
Technology (MoST) of Taiwan with Grant MoST 102-2119-M-007-004-MY3.
PMK acknowledges support through grant MoST 103-2119-M-001-009.
MP acknowledges the financial support of the OCEVU Labex (ANR-11-LABX-0060) and the A*MIDEX project 
(ANR-11-IDEX-0001-02) funded by the "Investissements d$'$Avenir" French government programme managed by the ANR.



\begin{deluxetable}{cccccccccccccccc}  \rotate
\tabletypesize{\scriptsize}   
\tablewidth{0pt}
\tablecaption{Analysis Summary
                      \label{table_quantities}}
\tablehead{
\multicolumn{5}{c}{Observation}
             & \colhead{}
             & \multicolumn{10}{c}{Analysis}\\
  \cline{1-5} \cline{7-15}
\colhead{Source / Region} & \colhead{$d$} & \colhead{$\lambda$}  & \colhead{$\theta$}  & \colhead{$\ell$} &
\colhead{}  & \colhead{$\mathcal{C}$} &
\colhead{$\delta_{max}$} &  \colhead{$\delta_{min}$} &
\colhead{$<|\delta|>$} &  \colhead{$std (|\delta|)$} &
\colhead{$\Sigma_{B,max}$} &  \colhead{$\Sigma_{B,min}$} &
\colhead{$<\Sigma_B>$} &  \colhead{$std (\Sigma_B)$} &
\colhead{Phase} \\
 \colhead{} & \colhead{(kpc)} & \colhead{($\mu$m)} &  \colhead{($\arcsec$)} & \colhead{(mpc)}  & \colhead{} &
\colhead{} & \colhead{(deg)} & \colhead{(deg)}
& \colhead{(deg)} & \colhead{(deg)}  &
 \colhead{} &  \colhead{} &  \colhead{} &  \colhead{} &  \colhead{} \\
 \colhead{} &  \colhead{(1)} &  \colhead{(2)} & \colhead{(3)} & \colhead{(4)} &  \colhead{} &  \colhead{(5)} &  \colhead{(6)} &
 \colhead{(7)} &  \colhead{(8)} &  \colhead{(9)} &
\colhead{(10)} & \colhead{(11)} & \colhead{(12)} & \colhead{(13)} & \colhead{(14)} 
}
\startdata

\multicolumn{1}{c}{SMA}\\ \cline{1-1}

G240&
4.7 &
870&
2.1 &
48 &
&
0.81 &
83   &
-83  &
24   &
23   &
1.62 (3.71) &
0.02  (0.02)&
0.35 (0.41) &
0.30 (0.52) &
III
\\

G34.4.0 &
1.57 &
870&
1.5 &
11 &
&
0.93 &
87 &
-61 &
28 &
23 &
1.16 (4.27) &
0.01 (0.001) &
0.26 (0.32) &
0.25 (0.54) &
III
\\

G34.4.1 &
1.57 &
870 &
1.5 &
11 &
&
0.42&
89 &
-89 &
34 &
25 &
10.6 (54.8) &
0.02 (0.007) &
0.87 (1.84) &
1.66 (6.55) &
IIA
\\

G35.2 &
2.19 &
870 &
1.5 &
16 &
&
0.63 &
90 &
-87 &
54 &
24 &
19.4 (194) &
0.02 (0.02) &
1.86 (5.11) &
3.22 (25.2) &
IIB
\\

NGC 6334A &
1.7 &
870 &
5.2 &
43 &
&
0.60 &
84 &
-84 &
56 &
21 &
8.96 (10.3) &
0.19 (0.19) &
1.91 (2.22) &
1.99 (2.53) &
IIB 
\\

NGC 6334I &
1.7 &
870 &
1.5 &
12 &
&
0.81 &
82 &
-90 &
50 &
23 &
10.1 (136) &
0.02 (0.003) &
1.14 (2.56) &
1.42 (14.0) &
IIB 
\\

NGC 6334IN &
1.7 &
870 &
2.4 &
20 &
&
0.81 &
79 &
-87 &
38 &
27 &
2.52 (3.93) &
0.04 (0.04) &
0.85 (0.91) &
0.52 (0.65) &
IIA 
\\

NGC 6334V &
1.7 &
870 &
2.9 &
24 &
&
0.78 &
79 &
-89 &
39 &
24 &
12.5 (49.5) &
0.02 (0.01) &
0.84 (1.57) &
1.86 (6.35) &
IIA
\\

I 18360 &
4.8 &
870 &
1.6 &
37 &
&
0.70 &
89 &
-86 &
41 &
28 &
3.95 (6.07) &
0.04 (0.04) &
0.58 (0.93) &
0.98 (1.66) &
IIB
\\

DR 21(OH) &
1.7 &
870 &
1.3 &
11 &
&
0.74 &
88 &
-87 &
40 &
26 &
7.04 (21.8) &
0.02 (0.007) &
0.99 (1.41) &
1.30 (2.73) &
-
\\

g5.89 $^{\ast}$ &
2 &
870 &
2.4 &
23 &
&
0.72 &
76 &
-84 &
35 &
26 &
2.88 (8.34) &
0.01 (0.008) &
0.55 (0.73) &
0.66 (1.41) &
- 
\\

Orion BN/KL  $^{\ast}$ &
0.48 &
870 &
2.8 &
6.4 &
&
0.70 &
89 &
-90 &
44 &
25 &
30.4 (136) &
0.02 (0.002) &
1.26 (4.72) &
3.99 (20.1) &
-
\\

W51 e2 - ext $^{\ast}$ &
7 &
870 &
0.7 &
24 &
&
0.95 &
33 &
-47 &
18 &
12 &
1.26 (1.35) &
0.02 (0.007) &
0.27 (0.30) &
0.30 (0.33) &
III
\\

W51 e2 - sub+comp &
7 &
870 &
2.0 &
68 &
&
0.92 &
77 &
-34 &
18 &
16 &
0.35 (0.35) &
0.06 (0.002) &
0.18 (0.17) &
0.09 (0.10) &
IIA/III
\\

W51 e8 - sub+comp &
7 &
870 &
2.0 &
68 &
&
0.76 &
85 &
-74 &
22 &
22 &
1.35 (1.35) &
0.12 (0.12) &
0.63 (0.63) &
0.31 (0.31) &
IIA/III
\\

W51 e2+e8 - sub+comp &
7 &
870 &
2.0 &
68 &
&
0.84 &
85 &
-74 &
21 &
20 &
2.0 (2.57) &
0.02 (0.02) &
0.44 (0.50) &
0.36 (0.51) &
IIA/III
\\

W51 North - ext $^{\ast}$ &
7       &
870       &
0.7        &
24        &
           &
0.89       & 
82         &
-81        & 
34       & 
23   &
2.48 (4.37) &
0.03 (0.005) &
0.64 (0.71) &
0.54 (0.79) &
IIA/III
\\

W51 North - sub$+$comp  &
7       &
870       &
2.0        &
68        &
           &
0.90       & 
70         &
-53        & 
27       & 
15   &
1.25 (1.25) &
0.03 (0.003) &
0.39 (0.36) &
0.31 (0.32) &
IIA
\\

NGC 1333 IRAS4A $^{\ast}$ &
0.3 &
870 &
1.2 &
1.7 &
&
0.82 &
84 &
-89 &
27 &
19 &
1.43 (4.27) &
0.01 (0.002) &
0.36 (0.43) &
0.29 (0.58) &
IIA/III
\\

IRAS 16293 A $^{\ast}$ &
0.12 &
870 &
2.5 &
1.5 &
&
0.86 &
90 &
-86 &
36 &
23 &
10.1 (177) &
0.02 (0.004) &
0.80 (1.48) &
1.08 (9.66) &
III
\\

IRAS 16293 B $^{\ast}$ &
0.12 &
870 &
2.5 &
1.5 &
&
0.57 &
89 &
-90 &
39 &
25 &
3.92 (10.1) &
0.01 (0.006) &
0.64 (0.83) &
0.60 (1.35) &
IIA/III
\\

G31.41 $^{\ast}$ &
7.9 &
870 &
0.9 &
34 &
&
0.92 &
73 &
-82 &
21 &
21 &
1.30 (3.83) &
0.01 (0.05) &
0.28 (0.35) &
0.23 (0.51) &
III \\

W3 (OH)$^{\ast}$  &
2.04 &
870 &
1.5 &
15 &
&
0.79 &
77	&
-87 &
35  &
22  &
2.50 (7.74) &
0.02 (0.001) &
0.54 (0.84) &
0.57 (1.64) &
IIA \\

&
&
&
&
&
&
&
&
&
&
&
&
&
&
&
\\

&
&
&
&
&
&
&
&
&
&
&
&
&
&
&
\\

\multicolumn{1}{c}{CSO}\\ \cline{1-1}

CO$+0.02$/ M$-0.02$  &
7.9   &
350   &
20    &
770   &
&
0.71 &
89   &
-88  &
48   &
25   &
13.9 (54.1) &
0.02 (0.02) &
1.69 (2.47) &
2.21 (5.97) &
I/-
\\

Mon R2  & 
0.95 &
350   &
20    &
92    &
&
0.78  &
67   &
-55  &
23  &
18   &
1.17 (1.17) &
0.01 (0.005) &
0.34 (0.32) &
0.29 (0.29) &
IIA

\\

M$+0.25+0.01$ &
7.9   &
350   &
20    &
770   &
&
0.61  &
88  &
-89 &
57  &
24   &
13.0 (38.6) &
0.01 (0.005) &
1.83 (2.41) &
2.88 (5.64) &
IIB
\\

NGC 2068 LBS10 &
0.4 &
350 &
20 &
39 &
&
0.62 &
77 &
-84 &
51 &
24 &
5.33 (5.33) &
0.02 (0.02) &
1.38 (1.38) &
1.67 (1.67) &
IIB

\\

NGC 2024 &
0.4 &
350  &
20  &
39 &
 &
0.55 &
85  &
-85 &
37 &
24 &
4.26 (7.17) &
0.05 (0.05) &
0.77 (0.93) &
0.83 (1.30) &
IIA

\\

NGC 2264 &
0.8 &
350 &
20 &
78 &
 &
0.94 &
42 &
-67 &
36 &
18 &
1.36 (1.36) &
0.03 (0.03) &
0.43 (0.43) &
0.46 (0.46) &
IIB

\\

NGC 6334A &
1.7 &
350 &
20 &
170 &
 &
0.84 &
86 &
-86 &
28 &
26 &
2.89 (7.27) &
0.06 (0.002) &
0.39 (0.59) &
0.54 (1.31) &
IIA
\\

G34.3$+0.2$ &
3.7 &
350 &
20 &
360 &
 &
0.87 &
89 &
-85 &
35 &
24 &
0.98 (17.8)&
0.01 (0.007)&
0.30 (0.82) &
0.22 (3.06) &
IIB 
 
\\

$\rho$ Oph &
0.139 &
350 &
20 &
14 &
 &
0.66 &
88 &
-78 &
27 &
24 &
1.02 (1.02) &
0.03 (0.03) &
0.50 (0.50) &
0.31 (0.31) &
IIA

\\

W49 A&
11.4 &
350 &
20 &
1110 &
&
0.86 &
77 &
-65 &
27 &
22 &
0.59 (1.02) &
0.02 (0.004) &
0.20 (0.22) &
0.18 (0.25) &
-

\\

GGD12&
1.7 &
350 &
20 &
170 &
&
0.98 &
56 &
-18 &
24 &
16 &
0.85 (0.85) &
0.02 (0.02) &
0.29 (0.29) &
0.29 (0.29) &
- 

\\

NGC 6334I&
1.7 &
350 &
20 &
170 &
&
0.63 &
88 &
-84 &
38 &
25 &
9.20 (34.9) &
0.01 (0.01) &
1.08 (1.71) &
1.46 (4.87) &
IIA
\\

M$+0.34+0.06$&
8 &
350 &
20 &
780 &
&
0.81 &
71 &
-83 &
58 &
26 &
5.06 (5.06) &
0.39 (0.39) &
1.60 (1.60) &
1.56 (1.56) &
IIB
\\

W33 C&
2.4 &
350 &
20 &
230 &
&
0.74 &
89 &
-75 &
40 &
27 &
1.09 (7.48) &
0.02 (0.02) &
0.31 (0.65) &
0.30 (1.59) &
-
\\

W33 A&
2.4 &
350 &
20 &
230 &
&
0.87 &
80 &
-75 &
36 &
22 &
1.17 (1.17) &
0.01 (0.003) &
0.28 (0.22) &
0.35 (0.33) &
IIA
\\

M 17 &
1.6 &
350 &
20 &
160 &
&
0.44 &
88 &
-90 &
34 &
26 &
7.18 (149) &
0.01 (0.01) &
0.75 (2.81) &
1.12 (15.9) &
IIA
\\

W3 &
1.95 &
350 &
20 &
190 &
&
0.91 &
78 &
-81 &
24 &
23 &
1.72 (3.01) &
0.01 (0.002) &
0.37 (0.48) &
0.42 (0.67) &
IIA
\\

OMC 1&
0.414 &
350 &
20 &
40 &
&
0.59 &
86 &
-90 &
29 &
20 &
11.2 (74.3) &
0.01 (0.001) &
0.74 (1.08) &
1.09 (4.50) &
IIA
\\

OMC 3&
0.414 &
350 &
20&
40 &
&
0.59 &
59 &
-89 &
25 &
20 &
1.57 (31.7) &
0.02 (0.02) &
0.68 (1.72) &
0.42 (5.68) &
IIA
\\

Sgr A$^{\ast}$ East &
8 &
350 &
20 &
780 &
 &
0.83 &
90 &
-90 &
47 &
30 &
14.1 (138) &
0.04 (0.04) &
2.08 (9.48) &
3.09 (30.0)  &
I/- \\

IRAS 05327&
0.414 &
350 &
20 &
40 &
&
0.43 &
81 &
-88 &
54 &
28 &
2.77 (2.77) &
0.11 (0.11) &
0.86 (0.86) &
0.92 (0.92) &
-
\\

Sgr B2&
8 &
350 &
20 &
780 &
&
0.70 &
90 &
-87 &
52 &
21 &
11.5 (63.0) &
0.02 (0.001) &
1.07 (1.47) &
1.46 (5.22) &
IIB
\\

Sgr B1&
8 &
350 &
20 &
780 &
&
0.76 &
90 &
-90 &
48 &
28 &
36.3 (204) &
0.06 (0.06) &
1.64 (8.33) &
4.94 (35.5) &
I / IIA
\\

DR21 &
3 &
350 &
20 &
290 &
&
0.58 &
86 &
-87 &
26 &
22 &
1.74 (3.61) &
0.01 (0.001)&
0.52 (0.50) &
0.37 (0.52) &
IIA
\\

Mon OB1 IRAS 12&
0.80 &
350 &
20 &
78 &
&
0.65 &
88 &
-88 &
42 &
29 &
5.56 (17.2) &
0.01 (0.01) &
0.78 (1.61) &
1.17 (3.74) &
IIB
\\

M$-0.13-0.08$&
8 &
350 &
20 &
780 &
&
0.71 &
90 &
-80 &
47 &
29 &
3.32 (123) &
0.07 (0.006) &
1.04 (6.73) &
0.98 (26.6) &
IIB 
\\

M$+0.07-0.08$&
8 &
350 &
20 &
780 &
&
0.89 &
38 &
-37 &
19 &
11 &
1.02 (1.02) &
0.01 (0.01) &
0.37 (0.37) &
0.38 (0.38) &
IIA
\\

\enddata
\tablecomments{Examples of $|\delta|$- and $\Sigma_B$-maps are shown in \citet{koch13} for CSO sources.
Values in parentheses for $\Sigma_{B,max}$, $\Sigma_{B,min}$, $\langle \Sigma_B \rangle$ and 
$std(\Sigma_B)$ represent values before removing statistical outliers with a $3\sigma$ clip.
Some values differ slightly from the ones in \citet{koch13} due to the way outliers were handled.
Sources with ``$^{\ast}$'' were previously observed outside of the SMA polarization legacy program. 
{\it ext, sub} and {\it comp} refer to the SMA extended, subcompact and compact array configurations.
Results for W51 e2, e8 and North are listed separately for different configurations as they are 
sampling different physical scales such as envelope or core \citep{tang12}.
Source coordinates of the original observations are listed in \citet{dotson10} and \citet{zhang14}.
}
\tablenotetext{(1)}{Source distance. Values are from \citet{genzel81} for W51 and Orion BN/KL, \citet{acord98}
                    for g5.89, \citet{racine70} for Mon R2 and from \citet{reid09} for CO$+0.02-0.02$ / M$-0.02-0.07$
                    and M$+0.25+0.01$. Further:  NGC 2068 LBS10 and NGC 2024 \citep{anthony82},
                               NGC 2264 \citep{park02},
                               NGC 6334A and 6334I \citep{russeil12},
			       G34.3$+0.2$ \citep{kuchar94},
			       $\rho$ Oph \citep{mamajek08},
			       W49 A \citep{gwinn92},
			       GGD12 \citep{rodriguez80},
			       sources around the galactic center, i.e., M$+0.34+0.06$, Sgr A$^{\ast}$ East,
                                 Sgr B2, Sgr B1, M$-0.13-0.08$, M$+0.07-0.08$ \citep{genzel00},
			       W33 C and W33 A \citep{immer13},			      
			       M17 \citep{povich07},
			       W3 \citep{xu06},
			       OMC 1, OMC 3 and IRAS 05327 \citep{menten07},			      			      
			       DR21 \citep{campbell82},
			       Mon OB1 IRAS 12 \citep{walker56},
			       IRAS 16293 \citep{loinard08}, G31.41 \citep{cesaroni94}, W3(OH) \citep{hachisuka06}.
			     }
\tablenotetext{(2)}{Observing wavelength.}
\tablenotetext{(3)}{Beam resolution. For elliptical beams (synthesized beams of the SMA) the geometrical mean is adopted.
                     For Hertz/CSO, the nominal beam size of $\sim$ $20\arcsec$ is listed.}  
\tablenotetext{(4)}{Physical size scales at the source distances for the resolutions $\theta$.}
\tablenotetext{(5)}{Correlation coefficient in the definition of Pearson's linear correlation coefficient
                    between magnetic field $P.A.$s and the intensity gradient $P.A.$s.}
\tablenotetext{(6)}{Maximum difference between magnetic field and intensity gradient orientation.}
\tablenotetext{(7)}{Minimum difference between magnetic field and intensity gradient orientation.}
\tablenotetext{(8)}{Mean absolute difference between magnetic field and intensity gradient orientations.}
\tablenotetext{(9)}{Standard deviation of absolute differences.}
\tablenotetext{(10)}{Maximum magnetic field significance.}
\tablenotetext{(11)}{Minimum magnetic field significance.}
\tablenotetext{(12)}{Mean magnetic field significance.}
\tablenotetext{(13)}{Standard deviation of magnetic field significance.}
\tablenotetext{(14)}{Assigned phase (I, IIA, IIB, III) according to the schematic scenario in Figure 11 in \citet{koch13}.
                     No phase is assigned to DR21(OH) (fragmenting core with multiple outflows \citep{girart13}),
g5.89 (probably in a later more evolved stage with expanding HII regions \citep{tang09a}) and Orion BN/KL 
(explosive outflow or pseudodisk \citep{tang10}).}
\end{deluxetable}

\begin{deluxetable}{ccccc}  
 \tabletypesize{\scriptsize}
\tablewidth{0pt}
\tablecaption{Source Type Properties
                      \label{table_source_type}}
\tablehead{
\multicolumn{1}{c}{Type}
& \multicolumn{1}{c}{Dust Morphology} & \multicolumn{1}{c}{Magnetic Field-Source Morphology} & \multicolumn{1}{c}{$|\delta|$ and $\Sigma_B$}
}
\startdata

 I          &   irregular   & irregular     & irregular  \\
IIA	    & elongated   & field mostly aligned with source minor axis   
                                 & systematic: overall small, larger towards and along central major axis  \\
IIB	    & elongated   & field mostly aligned with source major axis   
                                 & systematic: overall large, smaller towards and along central major axis  \\
III          & more circular  & field increasingly radial and / or hourglass-like  
                                 & systematic: overall small, additional azimuthal symmetries in $\pm\delta$ \\

\enddata
\tablecomments{Examples for IIA, IIB and III are shown in Figure \ref{figure_schematic} and \ref{figure_schematic_Sigma_B}.
Additional azimuthal symmetries appear in $\pm\delta$ (when giving it a sense of orientation) for 
type-III sources due to pinched magnetic field lines (Figure 6 in \citet{koch13}). 
Type-I sources are precursors where $|\delta|$- and $\Sigma_B$-properties are not yet clearly established.}
\end{deluxetable}

\end{document}